\documentclass{article}
\usepackage{textcomp} 

\title{Nonlinear Quantum Evolution Equations to Model  Irreversible
Adiabatic Relaxation with Maximal Entropy Production  \\ and Other
Nonunitary Processes}

\author{ Gian Paolo Beretta\\ Universit\`{a} di Brescia, via
Branze 38, Brescia, 25123 Italy \\ e-mail: beretta@ing.unibs.it
\\[2ex] }

\usepackage{amsmath}

\newcommand{\Hil}{{\mathcal H}}

\newcommand{\Boltz}{ k_{\rm\scriptscriptstyle B}}
\newcommand{\Tr}{{\rm Tr}}
\newcommand{\Ran}{{\rm Ran}}
\newcommand{\Ker}{{\rm Ker}}

\newcommand{\ddt}[1]{{\frac{{\rm d}#1}{{\rm d}t}}}
\newcommand{\Otimes}{{ \otimes }}
\newcommand{\g}{\ensuremath{\mathbf}}
\newcommand{\lt}{\left(}
\newcommand{\rt}{\right)}
\newcommand{\lqu}{\left[}
\newcommand{\rqu}{\right]}
\newcommand{\lat}{\left|}
\newcommand{\rat}{\right|}
\newcommand{\fr}{\frac}
\newcommand{\cov}[2]{{\langle\Delta #1\Delta #2\rangle}}
\newcommand{\covJ}[2]{{\langle(\Delta #1)^J(\Delta #2)^J\rangle}}
\newcommand{\covA}[2]{{\langle(\Delta #1)^A(\Delta #2)^A\rangle}}
\newcommand{\covB}[2]{{\langle(\Delta #1)^B(\Delta #2)^B\rangle}}
\newcommand{\mean}[1]{{\langle #1\rangle}}
\newcommand{\quot}[1]{{``#1''}}
\newcommand{\onehalf}{{\mbox{\textonehalf\,}}}
\newcommand{\onequarter}{{\mbox{\textonequarter\,}}}

\begin{document}

\maketitle

\begin{abstract}
We first discuss the geometrical construction and the main
mathematical features of the
maximum-entropy-production/steepest-entropy-ascent nonlinear
evolution equation proposed long ago by this author in the
framework of a fully quantum theory of irreversibility and
thermodynamics for a single isolated or adiabatic particle, qubit,
or qudit, and recently rediscovered by other authors. The
nonlinear equation generates a dynamical group, not just a
semigroup, providing a deterministic description of irreversible
conservative relaxation towards equilibrium from any
non-equilibrium density operator. It satisfies a very restrictive
stability requirement equivalent to the Hatsopoulos-Keenan
statement of the second law of thermodynamics. We then examine the
form of the evolution equation we proposed to describe
multipartite isolated or adiabatic systems. This hinges on novel
nonlinear projections defining local operators that we interpret
as ``local perceptions'' of the overall system's energy and
entropy. Each component particle contributes an independent local
tendency along the direction of steepest increase of the locally
perceived entropy at constant locally perceived energy. It
conserves both the locally-perceived energies and the overall
energy, and meets strong separability and non-signaling
conditions, even though the local evolutions are not independent
of existing correlations. We finally show how the geometrical
construction can readily lead to other thermodynamically relevant
models, such as of the nonunitary isoentropic evolution needed for
full extraction of a system's adiabatic availability.
\end{abstract}


\section{Introduction}

In this paper, we construct a  class of model evolution equations
(applicable not only to  open systems but also to closed isolated
systems) capable of describing---simultaneously and in competition
with the usual Hamiltonian unitary evolution---the natural
tendency of any initial nonequilibrium state to relax towards
canonical or partially-canonical thermodynamic equilibrium, i.e.,
capable of describing the irreversible tendency to evolve towards
the highest entropy state compatible with the instantaneous mean
values of the energy (and possibly other constants of the motion
and other constraints).

In this introduction, we briefly review some essential challenges
of a prevailing model of irreversibility (Section \ref{sec1-1}),
we briefly discuss the original motivation that lead us to develop
a quantum maximal entropy production formalism (Section
\ref{sec1-2}), we discuss the geometrical framework of our
derivation (Section \ref{sec1-3}), and we outline the structure of
the rest of the paper (Section \ref{sec1-4}).

\subsection{\label{sec1-1}Kossakowski-Sudarshan-Gorini-Lindblad quantum master
equation}

The prevailing model of irreversibility starts from unitary
dynamics but assumes that no system is truly isolated, so that
even an initial pure state becomes mixed due to increasing
system-environment entanglement.

The system-environment  entanglement builds up due to interactions
according to the standard Liouville-von Neumann unitary dynamics
of the overall system-environment composite. In this
phenomenological model  a system $A$ is assumed to be weakly
coupled with a reservoir $R$, so that they can exchange energy via
unitary evolution of the overall state $\rho_{AR}$. The reservoir
$R$ is modeled as a collection of a large number of quantum
systems (many degrees of freedom, e.g., the modes of the
electromagnetic field). Because of the weak coupling, the unitary
dynamics of $\rho_{AR}$ produces both an energy exchange and a
build up of correlations between the system and the reservoir.
However, justified only by some heuristic reasoning, a crucial
\emph{additional assumption} is injected in the derivation
(Markovian approximation): that correlations \emph{smear out}
rapidly enough so as to maintain $A$ and $R$ \emph{effectively}
decorrelated  not only initially, but at all times. One rationale
usually offered for this approximation is that when the reduced
density operator of $A$ is time averaged (coarse grained) over a
sufficiently long time interval, which is nevertheless still much
shorter than the system's time scale of interest, the average
correlations becomes negligible, and the averaged state
effectively factors at all times. The model is phenomenological
and basically charges the reservoir's complexity for the system's
losing quickly its memory of past interactions.\footnote{The
literature on the well known problem of accounting for the arrow
of time and the origin of irreversibility within Statistical
Mechanics and its time reversible underlying dynamics is too vast
to adequately review, see, e.g., \cite{Zeh}. Many alternatives to
time averaging have been proposed as rationales for coarse
graining (see, e.g., \cite{Uffink,Bassi,Albert}). We agree with a
Referee, that a common feature of all these attempts to justify
the continuing validity of the KSGL master equation, or
equivalents, is that they require additional, question begging
assumptions that are at odds with an overall unitary evolution.}
By tracing out all the reservoir's degrees of freedom, the overall
unitary dynamics gives rise (under such Markovian approximation)
to a system's reduced dynamics which is nonunitary, linear,
completely positive and generated by the celebrated
Kossakowski-Sudarshan-Gorini-Lindblad (KSGL) quantum master
equation \cite{Lindblad}
\begin{equation}
\label{Lindblad}
\ddt{\rho}=-\frac{i}{\hbar}[H,\rho]+\frac{1}{2}\sum_j\left(2
V_j^\dagger\rho  V_j -\{ V_j^\dagger V_j,\rho\}\right) \ ,
\end{equation}
where the $V_j$'s are some operators on $\Hil$ (each term within
the summation, often written in the alternative form $[V_j,\rho
V_j^\dagger]+ [V_j\rho, V_j^\dagger]$, is obviously traceless). It
has been used for  a number of successful models of dissipative
quantum dynamics of open subsystems. Operators $V_j$ in
(\ref{Lindblad}) are in general interpreted as creation and
annihilation, or transition operators. For example, by choosing
\cite{thesis}, $V_j=c_{rs}|r\rangle\langle s|$,  where $ c_{rs} $
are complex scalars and $ | s\rangle $ eigenvectors of the
Hamiltonian operator $H$, and defining the transition
probabilities $w_{r s}= c_{rs} c_{rs}^*  $, the KSGL equation
becomes
\begin{equation}
\label{Pauli1} \ddt{\rho} = - \, \frac{i}{\hbar} [ H , \, \rho ] +
\sum_{r s} w_{r s} \left( | s \rangle \langle r|\rho|r
\rangle\langle s | - \, \frac{1}{2} \{ | s \rangle\langle s | \, ,
\rho \} \right) \, ,
\end{equation}
or, equivalently, for the $ nm $-th matrix element of $\rho$ in
the $H$ representation,
\begin{equation}\label{Pauli2} \ddt{\rho_{nm}} =
 -\frac{i}{\hbar} \rho_{nm} ( E_n - E_m )+ \delta_{nm}
 \sum_r w_{n r} \rho_{r
r} - \rho_{nm} \frac{1}{2} \sum_r ( w_{r n} + w_{r m} ) \ ,
\end{equation}
which, for the $n$-th energy level occupation probability
$p_n=\rho_{nn}$, is the celebrated  Pauli master equation
\begin{equation}\label{Pauli3} \ddt{ p_n} = \sum_r w_{n r} p_r -
p_n \sum_r w_{r n} \  .
\end{equation}

In this widely accepted model, the assumption of \emph{erasure of
correlations}  is the sole mechanism responsible for \quot{entropy
generation} \cite{Gorini1976}, but the apparent contradiction with
the assumed underlying reversible unitary dynamics, i.e., the
Loschmidt paradox is still lurking behind. The situation is
parallel to what is needed to \quot{derive} the classical
Boltzmann equation from the underlying reversible
Hamilton-Liouville dynamics.

Philosophically, we find it hard to understand how diffusion of
mass, momentum, energy, and charge, could find their justification
in a \quot{loss of \emph{information} on the time scale of the
observer leading to rapid decoherence from the entanglement which
continuously builds up by weak coupling with environmental degrees
of freedom}. Is this the \emph{real} physical reason for the
\quot{universal tendency in nature to the dissipation of
mechanical energy} already recognized by Kelvin in 1852
\cite{Kelvin1852}? Do we have alternatives to understand and model
physical phenomena that are manifestly time asymmetric?

\subsection{\label{sec1-2}Locally maximal entropy production
dynamics as a nonlinear alternative to the KSGL model equations}

With these kind of motivation, thirty years ago we designed a
possible alternative based on the assumption, still to be
validated or invalidated, that irreversibility is a fundamental
microscopic dynamical feature and as such it must be built into
the fundamental laws of time evolution. Therefore, we constructed
a fundamental non-unitary extension of standard Schr\"{o}dinger
unitary dynamics not contradicting  any of the successful results
of pure-state quantum mechanics, and yet entailing the Second Law
as well as an objective entropy increase for mixed states of an
isolated system. We have shown in Refs.
\cite{thesis,ref41,ref42,ref44,Beretta2006group} that such an
approach is possible based on a steepest entropy ascent, i.e.,
maximal entropy generation, nonlinear and non-unitary equation of
motion which reduces to the Schr\"{o}dinger equation for pure
states. A 1985 Nature editorial \cite{Nature} defined this
approach \quot{an adventurous scheme which may end arguments about
the arrow of time}. Until now, however, the theory has been rather
ignored and neither validated nor invalidated experimentally.
Therefore (\quot{the proof of the pudding is in the eating}
\cite{Nature}), it remains just an interesting but little
acknowledged and pursued theoretical alternative to the standard
model. Some recent discussion about it, is found in
\cite{Challenge}. The challenge with this approach is to ascertain
if the intrinsic irreversibility it implies at the single particle
(local, microscopic) level is experimentally verifiable, or else
its mathematics must only be considered yet another
phenomenological tool, at the same level as the quantum Markovian
master equations which, as we have seen, are not free of their own
challenges.

The central conceptual difference between the proposed approach,
and the approaches based on attempting to derive the KSGL
equation, is that this approach\footnote{To our knowledge, the
first pioneering work where this assumption is made explicit and
used consistently to build a unified quantum theory of mechanics
and thermodynamics, is Ref. \cite{HG}. We thank a Referee for
suggesting the wording of this paragraph and the following
footnote.} regards a non-pure density matrix as representing a
real ontological object, the actual state of the world, and is not
understood as just an epistemic ignorance of which particular pure
state the world is `really' in.\footnote{If one assumes that the
`true' state of the world is actually a pure state, and that the
probabilities in a density matrix can only be a reflection of
uncertainty, then it is fairly easy to see that the evolution of
the density matrix must be linear: if $\rho_1 \rightarrow \rho'_1$
and $\rho_2 \rightarrow \rho'_2$, then a probabilistic mixture
$p_1\rho_1 +p_2\rho_2\rightarrow p_1\rho'_1 +p_2\rho'_2 $. This
linearity does not need to hold for our proposed evolution law as
a `real' density matrix $\rho$ is a distinct physical state, even
though numerically it may  be equal to $\rho = w_1\rho_1
+w_2\rho_2$ with $w_1 = p_1$ and $w_2 = p_2$, and is quite
different from a probabilistic mixture arising through ignorance
or uncertainty. Even if  $\rho_1 \rightarrow \rho'_1$ and $\rho_2
\rightarrow \rho'_2$, it does not follow that $\rho = w_1\rho_1
+w_2\rho_2¨ \rightarrow w_1\rho'_1 +w_2\rho'_2 $. This conceptual
difference is at the heart of our original approach. The physical
reality attributed to the density matrix also legitimates treating
the entropy $-\Boltz\Tr(\rho\ln\rho)$ as a `real' physical
quantity, in the manner of energy or mass, and not as an
expression of information or uncertainty about a probability
distribution.}

Therefore, we avoid the (unnecessary) severe restrictions imposed
by linearity on the evolution equation, and we open up our
attention to evolution equations nonlinear in the density operator
$\rho$. It may at first appear natural to maintain the
Kossakowski-Lindblad form (\ref{Lindblad}) and simply assume that
operators $V_j$ are functions of $\rho$. This is true only in part
for the evolution equation we will construct. Indeed, our
hermitian operator $\Delta M$ in our Eq. (\ref{rhodot}) below, can
always be written as $-\sum_j V_j^\dagger(\rho) V_j(\rho)$ and
therefore our anticommutator term may be viewed as a
generalization of the corresponding term in (\ref{Lindblad}).

However, our geometrically motivated construction based on the
square-root of the density operator effectively suppresses the
term corresponding to $\sum_j V_j^\dagger\rho V_j$ in
(\ref{Lindblad}). The reason we find  this suppression desirable
is the following. Due to the terms $V_j^\dagger\rho V_j$
(\ref{Lindblad}), whenever $\rho$ is singular, its zero
eigenvalues may change at a finite rate. This can be seen clearly
from (\ref{Pauli3}), by which ${\rm d}p_n/{\rm d}t$ is finite
whenever there is a nonzero transition probability $w_{nr}$ from
some other populated level ($p_r\ne 0$), regardless of whether
$p_n$ is zero or not. When this occurs, for one instant in time
the rate of entropy change is infinite, as seen clearly from the
expression of the rate of entropy change implied by
(\ref{Lindblad}),
\begin{equation}\label{PauliS1} \ddt{\mean{S}} =\Boltz \sum_j
\Tr ( V_j^{\dagger}
 V_j \rho \ln
\rho - V_j^{\dagger} \rho V_j \ln \rho )= \Boltz \sum_{j r n}
 ( V_j )^*_{n r}  ( V_j )_{n r} ( \rho_r - \rho_n ) \ln \rho_r \ ,
\end{equation}
where $\rho_r$ denotes the $r$-th eigenvalue of $\rho$ and $( V_j
)_{n r}$ the matrix elements of $V_j$ in the $\rho$
representation.

We may argue that an infinite rate of entropy change  can  be
tolerated, because it would last only for one instant in time. But
the fact that zero eigenvalues of $\rho$ in general  would not
remain zero (or close to zero) for longer than one instant in
time, to us is an unphysical feature, at least because it is in
contrast with a wealth of successful models of physical systems in
which great simplification is achieved  by limiting our attention
to a restricted subset of relevant eigenstates (forming a subspace
of $\Hil$ that we call the effective Hilbert space of the system
\cite{MPLA}). Such common practice models yield extremely good
results, which, being reproducible, ought to be relatively robust
with respect to inclusion in the model of other less relevant
eigenstates. In fact, such added eigenstates, when initially
unpopulated, are irrelevant if they remain unpopulated (or very
little populated) for long times, so that neglecting their
existence should introduce very little error. The terms
$V_j^\dagger\rho V_j$, instead, would rapidly populate such
irrelevant unpopulated eigenstates and void the validity of our so
successful simple models. Of course,  we may deliberately overlook
this instability problem by making highly ad-hoc assumptions,
e.g., by forcing the $V_j$'s to be such that $( V_j )_{n r}=0$
whenever either $\rho_n=0$ or $\rho_r=0$. But, in this case, we
can no longer claim true linearity with respect to $\rho$.

Another important general physical reason why we find it
advantageous that our construction excludes KSGL terms that
generate nonzero rates of change of the zero eigenvalues of
$\rho$, is that such terms are construed so as to preserve the
positivity of $\rho$ in forward time, but in general they do not
maintain it in backward time. Such mathematical irreversibility of
the  Cauchy problem is often accepted, presented, and justified as
a natural counterpart of physical irreversibility. However, we
already noted in \cite{MPLA} that it is more related to a
principle of causality than to physical irreversibility. The
strongest form of a general non-relativistic principle of
causality---a keystone of traditional physical thought---requires
that future states of a system should unfold deterministically
from initial states along smooth unique trajectories in the state
domain, defined for all times, future as well as past. Accepting
mathematical irreversibility of the model dynamics, would imply
giving up such causality requirement. The foundational virtue of
our dynamical group is in its very existence, which shows a simple
conceivable alternative whereby we are not compelled to cope with
such a major conceptual loss. Regardless of these important but
highly controversial foundational implications, we have shown in
\cite{PRE} that our Eq. (\ref{rhodot}) can effectively describe
relaxation within an isolated system, and yet it is mathematically
reversible, in the sense that it features existence and uniqueness
of well-defined solutions both in forward and backward time.

Eq. (\ref{rhodot}) describes physically irreversible time
evolutions, in the sense that the physical property described by
the entropy functional $-\Boltz\Tr(\rho\ln\rho)$ is a strictly
increasing function of time for all states except  the very
restricted set of equilibrium states and limit cycles defined by
Eq. (\ref{nondissH}) below.

Similarly to our presentation of MEPP dynamics in a general
probabilistic but non-quantal framework in \cite{ASME,Entropy}, in
this paper we focus on the mathematical features and the potential
phenomenological applications of a quantal MEPP dynamical
equation. We emphasize that the formalism has an intrinsic
mathematical validity \emph{per se} as a mere phenomenological
tool. This may be useful also for those who remain understandably
skeptical about the cited adventurous scheme of our original
attempt to unify mechanics and thermodynamics, whereby, again, we
proposed a resolution of the long-standing dilemma about the arrow
of time based on building the Hatsopoulos-Keenan statement of the
second law directly into the dynamical postulate of quantum
theory.

On the other hand, if proved valid at the fundamental level as
envisioned in its original framework, our nonlinear dynamical law
would imply the incompleteness of unitary pure-state zero-entropy
quantum mechanics and the need to broaden it as suggested in
\cite{HG,MPLA2}. In such context, our microscopic dynamical theory
might also be seen to accomplish the program sought for with
limited  success in the 1980's by the Prigogine school
\cite{Jones1982,Prigogine1983}, namely, to build a mathematical
theory of microscopic irreversibility (the question:
\quot{\emph{minimal} entropy production or \emph{maximal} entropy
production?} is an interesting one, and, at least mathematically,
is clarified in \cite{Dewar2005,Struchtrup} where it is shown that
\emph{maximal entropy production in general}, implies
\emph{minimal} entropy production at some constrained stationary
states).

\subsection{\label{sec1-3}Maximal Entropy Production Path in a
Maximum Entropy Landscape}

The determination of a density operator of maximum entropy subject
to a set of linear constraints has applications in many areas of
quantum physics, chemistry,  information, and probability theories
\cite{ref11,ref12,ref13,Keck,KB}. The maximum entropy density
operator typically represents a thermodynamic equilibrium state or
a constrained-equilibrium state of the system under study.

Having set aside the cited implications on conceptual and physical
quantum foundations, this paper focuses on the geometrical
construction of our MEPP nonlinear quantum master equation,
presented as the mathematical generalization of the maximum
entropy problem to the nonequilibrium domain, by discussing a
general rate equation for the description of smooth constrained
relaxation of arbitrary non-equilibrium density operators towards
maximum entropy. The nonlinear rate equation keeps the constraints
constant at their initial values and increases the entropy until
an unstable or stable maximum-entropy equilibrium state is
approached. The unstable equilibrium density operators are those
with at least one zero eigenvalue and all others canonically
distributed (see Eq. (\ref{nondissH}) below). The rate equation is
also consistent with an Onsager reciprocity theorem interestingly
extended to the entire non-equilibrium domain.

Geometrically, it has a clear representation in square-root
density operator state space. Every trajectory unfolds along a
path of steepest entropy ascent compatible with the constraints
(constrained geodesics). For an isolated system, the constraints
represent constants of the motion. For more general quantum
thermodynamics modeling, such as for rate-controlled constrained
equilibrium modeling of chemical kinetics \cite{Keck,KB}, the
constraints may be assigned a specified time-dependence.

The well-known maximum entropy problem which sets our context
(landscape, to use the terminology of nonlinear optimization)  is
that of seeking a density operator $\rho$ whose entropy $ S(\rho)
= -\Boltz \Tr\rho\ln\rho$  is maximal subject to given magnitudes
$ \langle A_k \rangle $ of one or more linear constraints $
\Tr\rho A_k = \langle A_k \rangle$ for $ k = 0, 1, \ldots, n $
where $ A_k $ is the hermitian operator associated with the $
k$-th constrained observable. We assume  the first constraint to
be the normalization condition, so that $ A_0 = I $ and $ \langle
A_0 \rangle = 1 $. Moreover, as suitable to model a canonical
isolated system, below we will assume for simplicity a single
nontrivial constrained observable, the energy, represented by the
Hamiltonian operator $A_1=H$.

The maximizing density operator $ \rho^* $ can be written as
$\rho^* =  \exp \lt - \sum_{k = 1}^{n} \lambda_k \, A_{k } \rt/Q$
with $Q = \Tr \exp \lt - \sum_{k = 1}^{n} \lambda_k \, A_{k } \rt
$ where the Lagrange multipliers $ \lambda_k $ are determined by
the values $ \langle A_k \rangle$ of the constraints.

In this landscape, we wish to consider the following general
problem in the non-equilibrium domain. We seek a time-dependent
density operator, namely, an operator function (one-parameter
family) $ \rho(t) $, whose zero eigenvalues remain zero at all
times and whose entropy $ S \lt
 \rho(t) \rt $ is maximally increasing with time $t$, i.e.,
 \begin {equation}\label{Snondecrease} \mbox{max }  -\Boltz \ddt{} \Tr\rho(t)\ln\rho(t) \ge  0
 \ \mbox{  subject to  }\   \Tr\rho(t) A_{k}
 =\mbox{const \ and\ } \ddt{\ell(t)}=\mbox{const,}
 \end{equation}
where $\ell(t)$ is a properly formulated measure of length of a
trajectory in density operator state space.  In time interval
$dt$, among all the possible trajectories in state space that have
length $d\ell$, the system selects that which yields the maximal
increase in the value of the entropy functional, i.e., the path of
steepest entropy ascent. This models a most irreversible quantum
evolution towards maximum entropy. It is a realization at the
(fundamental?) quantum dynamical level of the \quot{principle of
maximal entropy generation} \cite{Frontiers,LecNotes}. The
empirical validity of such a principle at the phenomenological
level has been recently affirmed (explicitly or implicitly) by
various authors in different fields and frameworks (see, e.g.,
\cite{Dewar2005,Ozawa2003,Grmela,Caticha2002,Chung,Martyusheva}).

Though overlooked even in recent reviews \cite{Martyusheva}, our
nonlinear dynamics is one of the earliest instances and
implementations of the ansatz which today goes under the name of
maximum entropy production principle (MEPP). A reason for the
oversight may have been that starting in 1984
\cite{Frontiers,LecNotes} we classified our approach as `steepest
entropy ascent' dynamics, rather than MEPP, to emphasize that the
qualifying and unifying feature of this dynamical principle is the
direction of maximal entropy increase rather than the rate at
which a nonequilibrium state is attracted in such direction.

The formalism presented here has mathematical features of great
generality, and is presented in a form readily adaptable to
different applications. It was originally \quot{designed} by the
author in 1981 \cite{thesis} and subsequently developed
\cite{ref41,ref42,ref44,PRE,Frontiers,LecNotes} to obtain an
equation of motion for a quantum theoretical unification of
mechanics and thermodynamics \cite{Nature,HG}.

Recently, the original equation has been partially rediscovered in
the same context \cite{Gheorghiu}. The idea of
steepest-entropy-ascent time evolution of a probability
distribution has also been recently rediscovered in
\cite{Lemanska} but with important differences we discuss in
\cite{PRE}. Because of its intriguing general features, we
suggested long ago \cite{ASME} that the formalism maintains its
appeal even when abstracted from its original physics purpose
because it provides a powerful mathematical tool for
phenomenological modeling applications. It has  been recently
rediscovered also in such a broader maximum entropy formalism,
probabilistic context \cite{Caticha2001}.

\subsection{\label{sec1-4}Outline of the Paper}

In Section \ref{sec2} we present the geometrical reasoning that
leads to the construction of our main equation, Eq. \ref{rhodot},
for a single particle system. Important to this development (as
well as  the rest of the paper) is the material in the Appendix,
which reviews some well known but little used geometrical notions,
and sets the notation of our derivations.

In Section \ref{sec3} we outline the main features and theorems of
Eq. \ref{rhodot}.

In Section \ref{Composite} we discuss the generalization of our
dynamics  to composite systems, which is nontrivial in view of the
nonlinearity of the steepest entropy approach.

Finally, in Section \ref{Further} we discuss a further
generalization of the foregoing non-equilibrium problem whereby
the magnitudes $ \langle A_k \rangle $ of the constraints and the
entropy rate of change may be assigned definite or interrelated
time-dependences. This may become useful in the framework of
quantum thermodynamics modeling of a non-work interaction, by
which we mean \cite{GB2005} an interaction where in addition to
energy exchange between the interacting systems, there is also
entropy exchange.

\section{\label{sec2}Geometrical construction of a single-particle
MEPP quantum dynamics}

\subsection{\label{sec2-1}Reformulation in terms of square-root
density operators}

Because we seek a well-defined time evolution equation for the
density operator, we must enforce at all times  the positive
semi-definiteness and hermiticity constraints, i.e., $
\rho(t)^\dagger=\rho(t) \geq 0 $. To this end it is convenient to
change variables and represent quantum states by means of the
square-root density operator $\gamma$ defined as follows
\begin{equation} \gamma =U\sqrt{\rho}\ , \qquad U^\dagger=U^{-1} \
, \qquad \rho=\gamma^\dagger\gamma \ ,
\end{equation} where $\sqrt{\rho}$ is the positive square root of
$\rho$, and $U$ an arbitrary unitary operator that in the end will
turn out to be irrelevant, much like  phase factors in usual
quantum mechanics. In the original derivations we assumed $U=I$,
but as suggested in \cite{Gheorghiu} the introduction of $U$ has
some formal advantage.

Notice that were it only to cope with the positive semi-definite
constraint, we could choose as new 'variable' any function of
$\rho$ whose inverse is even. The main reason for choosing $
\sqrt{ \rho }$ is geometrical and part of the steepest entropy
ascent assumption. We note here that, like done  in Ref.
\cite{Lemanska} in a non quantal context, we could derive a
steepest entropy ascent dynamics without switching to a square
root representation, but in such case the entropy gradient would
not be well-defined on the entire domain, it would diverge
whenever one of the eigenvalues of $\rho$ is zero, and as a result
the dynamics would exhibit unphysical infinite-rate effects and
would not conserve positivity when solved backwards in time.

In order to introduce the geometrical notion of steepest entropy
ascent, we need to define what we mean by \quot{distance} between
two density operators and by \quot{length} of a portion of
trajectory in state space, i.e., a one parameter family of density
operators, a time evolution. The proper unique natural
 metric for this purpose is known in statistics as the Fisher-Rao
metric (see e.g. \cite{Wootters,ref6,Caves}).  For a one-parameter
family of discrete distributions, $\g{p}(t)$, where $t$ is the
parameter, the distance between distributions $\g{p}(t+dt)$ and
$\g{p}(t)$ is
\begin{eqnarray}
d\ell &=& \frac{1}{2}\sqrt{\sum_i p_i\left(\frac{d\ln
p_i}{dt}\right)^2}\,dt =\frac{1}{2}\sqrt{\sum_i
\frac{1}{p_i}\left(\frac{d p_i}{dt}\right)^2}\,dt =\sqrt{\sum_i
\left(\frac{d\sqrt{p_i}}{dt}\right)^2}\,dt\nonumber \\
&=&\sqrt{\sum_i \left(\dot x_i\right)^2}\,dt=\sqrt{\g{\dot
x}\cdot\g{\dot x}}\,dt \ .
\end{eqnarray}
Thus,  square-root probabilities $x_i=\sqrt{p_i}$ are the most
natural variables in that: \begin{itemize} \item the space becomes
the unit sphere, $\g{x}\cdot\g{x}=1 $ $(\sum_i p_i=1)$; \item the
Fisher-Rao metric simplifies to $d\ell=\sqrt{\g{\dot
x}\cdot\g{\dot x}}\,dt$, or equivalently $d\ell^2=d\g{x}\cdot
d\g{x}$;
\item the distance between any two distributions is the
angle $d(\g{x}_1,\g{x}_2)= \cos^{-1}(\g{x}_1\cdot\g{x}_2)$.
\end{itemize}

We therefore conveniently rewrite the density operator formalism
in terms of the square-root-density operator representation of
states. To do so, we equip the space of linear (not necessarily
hermitian) operators on $\Hil$ with  the \emph{real} scalar
product\footnote{Note that this \emph{real} inner product on the
vector space of linear operators does satisfy the necessary rules,
including of course that $X\cdot Y=Y\cdot X$, $X\cdot X\ge 0$, and
$X\cdot X =0$ iff $X=0$. It clearly differs from the more usual
\emph{complex} inner product $\Tr(X^\dagger Y)$.}
\begin{equation}\label{scalar} X\cdot Y = \onehalf\Tr (X^\dagger
Y+Y^\dagger X) \ .\end{equation} The state space becomes the unit
sphere $\gamma\cdot\gamma=1$ ($=\Tr\rho$ with
$\rho=\gamma^\dagger\gamma$ automatically positive semidefinite).
On the state space we therefore adopt the Fisher-Rao type of
metric $d(\gamma_1,\gamma_2)= \cos^{-1}(\gamma_1\cdot \gamma_2)$,
so that along a time dependent trajectory
\begin{equation}\label{ell}d\ell =
2\sqrt{\dot\gamma\cdot\dot\gamma}\,dt \ .\end{equation}

\subsection{Notation. Gradients of the Energy and Entropy Functionals}

For simplicity, in addition to the normalization constraint, here
we will assume a single additional constraint, namely, energy
conservation,\footnote{If $H$ depends on a set of time dependent
parameters $\lambda_1$, $\lambda_2$,\dots, $\lambda_K$, the system
is \emph{adiabatic} and we usually interpret $\langle
dH/dt\rangle=\sum_k\langle \partial
H/\partial\lambda_k\rangle\dot\lambda_k $ as the rate of work
exchange between the system and a set of $K$ \quot{work elements}
\cite{HG} mechanically coupled with the system through the
variation of these parameters. Such an adiabatic system undergoes
what in \cite{GB2005} we call a \quot{weight process}, and the
energy balance equation (energy conservation) reads $ d\langle
H\rangle/dt=\langle dH/dt\rangle$ or, in the notation we introduce
in this section, $H'\cdot\dot\gamma=0$. } with associated
Hamiltonian $H$ hermitian on $\Hil$. The extension to more
constraints is straightforward \cite{ArXiv1} in view of the
formalism in the Appendix. In terms of the square-root density
operator $\gamma$, the functionals representing the mean values,
their time rates of change along a time dependent trajectory
$\gamma(t)$, and the dispersions and covariance of the energy and
the entropy, are conveniently rewritten introducing the following
notation\footnote{It hinges on the inner product defined by
(\ref{scalar}). The logic and some details are as follows. For any
hermitian $A$, we define $A'=2\gamma A$ so that $\langle A\rangle
= \Tr(\rho A)=\onehalf A'\cdot\gamma$ (clearly, in general $A'$ is
not hermitian). When the $'$ operation is applied on $\Delta
A=A-\langle A\rangle I$, we obtain $(\Delta A)'=2\gamma \Delta A$
and, in general, $(\Delta A)'\cdot\gamma=0$. Next, because
$\rho=\gamma^\dagger\gamma$ and $\Tr\rho=\gamma\cdot\gamma=1$, we
have $\Tr\dot\rho=\dot\gamma\cdot\gamma+
\gamma\cdot\dot\gamma=2\gamma\cdot\dot\gamma=0$, $\dot\gamma A
\cdot\gamma= \gamma A\cdot\dot\gamma$ and, therefore,
$\Tr(\dot\rho
A)=\Tr[(\dot\gamma^\dagger\gamma+\gamma^\dagger\dot\gamma)A]=
\onehalf\Tr[\dot\gamma^\dagger(2\gamma A)+(2\gamma
A)^\dagger\dot\gamma]=A'\cdot\dot\gamma=(\Delta A)'\cdot\dot\gamma
$. In general, therefore, $d \langle A\rangle/dt- \langle dA/dt
\rangle= \Tr(\dot\rho A)=A'\cdot\dot\gamma$. For a time
independent Hamiltonian $H$, $d \langle H\rangle/dt=
H'\cdot\dot\gamma$. Moreover, because $\Tr(\rho\dot S)=0 $ (proof
in the next footnote), we have $d \langle S\rangle/dt=
S'\cdot\dot\gamma $ in spite of $S=-\Boltz P_{\Ran\rho} \ln\rho$
being time dependent. Finally, for any hermitian pair $A$ and $B$,
we have the identity $\gamma\Delta A\cdot\gamma\Delta
B=\onehalf(\langle\Delta A\Delta B\rangle+\langle\Delta B\Delta
A\rangle)$. Because $\langle\Delta A\Delta B\rangle- \langle\Delta
B\Delta A\rangle=\Tr(\rho [A,B])$, in general $\langle\Delta
A\Delta B\rangle\ne \langle\Delta B\Delta A\rangle$ unless $A$ and
$B$ commute or one of them commutes with $\rho$. }$^,$\footnote{To
show that $\Tr(\rho\dot S)=0 $, we let $B=P_{\Ran\rho}$ and use
the identity $B\dot B B=0$ which follows from $B^2=B$, $\dot B
B+B\dot B=\dot B$, $B\dot B B+B\dot B=B\dot B$. Let $P_\alpha$ be
the one-dimensional projectors $|\alpha\rangle\langle \alpha|$
onto the eigenvectors of $\rho$ with non-zero eigenvalues
$p_\alpha$ (repeated if degenerate). Then, $P_\alpha
P_\beta=\delta_{\alpha\beta}P_\alpha$, $B=\sum_\beta P_\beta$,
$P_\alpha B=B P_\alpha =P_\alpha$, $\rho=\sum_\alpha P_\alpha
p_\alpha$, $\Tr\rho=\sum_\alpha p_\alpha=1$, $\sum_\alpha \dot
p_\alpha=0$, $S=-\Boltz \sum_\beta P_\beta\ln p_\beta$, always
well defined because the sum is restricted to the nonzero
$p_\beta$'s, $\dot S=-\Boltz \sum_\beta P_\beta\dot
p_\beta/p_\beta-\Boltz \sum_\beta \dot P_\beta\ln p_\beta$, and
finally $\Tr(\rho\dot S)=-\Boltz \sum_\beta \dot p_\beta-\Boltz
\sum_\alpha\sum_\beta \Tr(P_\alpha\dot P_\beta) p_\alpha\ln
p_\beta= -\Boltz \sum_\alpha \Tr(P_\alpha\dot P_\alpha)
p_\alpha\ln p_\alpha=0$ because $\Tr(P_\alpha\dot
P_\beta)=\delta_{\alpha\beta}\Tr(P_\alpha\dot P_\alpha)$ and
$0=\Tr(B\dot B BP_\alpha)=\Tr(B\dot B P_\alpha)=\Tr(P_\alpha\dot B
)=\sum_\beta\Tr(P_\alpha\dot P_\beta)=\Tr(P_\alpha\dot
P_\alpha)$.}
\begin{align}
 H'&= 2\gamma H  \mbox{, so that } \Tr\rho H
=\langle H\rangle= \onehalf H'\cdot\gamma ; \\ S'&= 2\gamma S
\mbox{, with } S=-\Boltz P_{\Ran\rho}\ln\rho = -\Boltz
\ln(\rho+P_{\Ker\rho} ) \mbox{, so that}\\ \langle S\rangle
&=-\Boltz\Tr\rho \ln\rho = \onehalf S'\cdot\gamma ;\\ (\Delta
H)'&= 2\gamma \Delta H \mbox{, with } \Delta H=H-\langle H\rangle
I \mbox{, so that } (\Delta H)'\cdot\gamma=0 ;\\ (\Delta S)'&=
2\gamma \Delta S \mbox{, with } \Delta S=S-\langle S\rangle I
\mbox{, so that } (\Delta S)'\cdot\gamma=0 ;\\ \dot\rho
&=\dot\gamma^\dagger\gamma+\gamma^\dagger\dot\gamma \mbox{, since
} \rho=\gamma^\dagger\gamma;\\
\Tr\dot\rho&=\dot\gamma\cdot\gamma+\gamma\cdot\dot\gamma=
2\gamma\cdot\dot\gamma=0 \mbox{, since }
\Tr\rho=\gamma\cdot\gamma=1;\label{unitrate}\\ d\langle
H\rangle/dt&-\langle dH/dt\rangle =\Tr\dot\rho H =\onehalf
(\dot\gamma\cdot H'+\gamma\cdot\dot H')= H'\cdot \dot\gamma;
&\label{energyrate}\\ d\langle S\rangle/dt&=d(-\Boltz\Tr\rho
\ln\rho)/dt = S'\cdot\dot\gamma;  & \label{entropyrate} \\
\cov{H}{H}&=\Tr\rho(\Delta H)^2=\gamma \Delta H\cdot \gamma \Delta
H=\onequarter (\Delta H)'\cdot(\Delta H)';\\
\cov{S}{S}&=\Tr\rho(\Delta S)^2=\gamma \Delta S\cdot \gamma \Delta
S=\onequarter (\Delta S)'\cdot(\Delta S)';\\
\cov{S}{H}&=\cov{H}{S}=\onehalf \Tr\rho\{\Delta S,\Delta
H\}=\gamma \Delta S\cdot \gamma \Delta H=\onequarter (\Delta
S)'\cdot(\Delta H)';
\end{align}
where $P_{\Ran\rho}=I-P_{\Ker\rho}$ is the projector onto the
range of $\rho$ (the subspace given by the eigenvectors of $\rho$
with non-zero eigenvalues). Because of Eqs. (\ref{unitrate}),
(\ref{energyrate}) and (\ref{entropyrate}) we call the operators
$2\gamma$, $H'$ and $S'$ the normalization gradient, energy
gradient and entropy gradient operators, respectively. In the same
sense, $(\Delta H)'$ and $(\Delta S)'$ are the gradients of the
null functionals $\Tr\rho(\Delta H)'$ and $\Tr\rho(\Delta S)'$.

It is noteworthy that a dual set of definitions may be constructed
using $\tilde\gamma=\sqrt{\rho}U=\gamma^\dagger$,
$\rho=\tilde\gamma\tilde\gamma^\dagger$, $\tilde
H'=2\tilde\gamma^\dagger H$, $d\langle H\rangle/dt-\langle
dH/dt\rangle =\tilde H'\cdot \tilde\gamma^\dagger$, and so on,
leading however to the same results.

So, thinking geometrically, $\gamma(t)$ is a parameterized path in
the (square root) density matrix space and $\dot\gamma(t)$ gives
the tangent vector to this path. If the hermitian operator $A$ is
not varying directly with time, $A'(t)\cdot \dot\gamma(t)$ gives
the rate of change of its mean value $\langle A\rangle (t)$ as one
follows the given path. To maximize the rate of change of the mean
value, the tangent vector $\dot\gamma(t)$ should be parallel to
$A'(t)$, whereas to hold the mean value constant, it should be
perpendicular. From this follows the interpretation of $2\gamma$,
$H'$ and $S'$ as the normalization, energy, and entropy gradients,
respectively. For a trajectory $\gamma(t)$ to conserve
normalization and energy, the tangent vector $\dot\gamma(t)$ must
be always  perpendicular to $\gamma$ and $H'$. This holds
automatically for unitary evolutions, for which $\dot\gamma(t)$ is
perpendicular also to $S'$ (see Section \ref{sec2-3}). For a more
general evolution, to maximize the generation of entropy,
$\dot\gamma(t)$ should be parallel to $S'$, but in general this is
in conflict with the conservation of normalization and energy. So,
in Section \ref{sec2-4} we take $\dot\gamma(t)$ to be parallel to
the component of $S'$ perpendicular to both $\gamma$ and $H'$,
i.e., in the direction of greatest entropy increase consistent
with the constraints of normalization and energy conservation.

\subsection{\label{sec2-3}Hamiltonian Evolution and Time-Energy
Uncertainty Relations}

Eqs. (\ref{unitrate}) and (\ref{energyrate}) imply that  $\gamma$
remains unit norm and $ d\langle H\rangle/dt=\langle dH/dt\rangle$
(energy conservation)\footnote{In particular, $\langle H\rangle$
is time invariant if $H$ is time independent (isolated system).}
when $\dot\gamma$ is orthogonal to both $\gamma$ and the energy
gradient vector $H'$.

This is the case for purely Hamiltonian evolution, whereby
\begin{equation}\label{gammadotH} \dot\gamma_H=i\gamma \Delta H/\hbar
 \quad \Rightarrow \quad  \dot\rho_H=-i[H,\rho]/\hbar\  .\end{equation}
 Note that $\dot\gamma_H$ is
everywhere orthogonal also to $S'$ and hence also the entropy is
time invariant. It is actually so in a very special way, as each
eigenvalue of $\rho$ is time invariant under unitary evolution.
Note also that, since $\gamma \Delta H\cdot \gamma \Delta
H=\cov{H}{H}= \hbar^2\dot\gamma_H\cdot\dot\gamma_H$,
 the Fischer-Rao metric
$d\ell = 2\sqrt{\dot\gamma_H\cdot\dot\gamma_H}\,dt = dt/\tau_H$
 defines an intrinsic Hamiltonian time $\tau_H$ such that
\begin{equation} \cov{H}{H}\,\tau^2_H=\hbar^2/4 \ .\end{equation}
This can be interpreted as the shortest characteristic time of
unitary evolution, because indeeed the standard
(Mandelstam-Tamm-Messiah \cite{Messiah}) time-energy uncertainty
relation  can be stated as follows \cite{ArXiv1,ArXivTE},
\begin{equation}\label{timeenergy} \tau^2_{F\!H}=\cov{F}{F}/|d\langle F
\rangle/dt|^2\ge \tau^2_H  \quad \Rightarrow \quad \cov{H}{H}\,
\tau^2_{F\!H} \ge \hbar^2/4 \ ,\end{equation} where $F$ is any
hermitian operator and $ \tau_{F\!H}$ is the characteristic time
of change of its mean value $\langle F \rangle$ under Hamiltonian
evolution. A well-known implication of (\ref{timeenergy}) is that
the mean value $\langle F \rangle$ of any observable cannot change
at a rate exceeding $2\sqrt{\cov{F}{F}\cov{H}{H}}/\hbar$.

\subsection{\label{sec2-4}Time Evolution Along the Path of Steepest
Entropy Ascent}

Instead of pure Hamiltonian evolution, let us assume
\begin{equation}\label{gammadot}\dot\gamma= \dot\gamma_H+
\dot\gamma_D\ ,\end{equation} with $\dot\gamma_D$ in the direction
of steepest entropy ascent compatible with the constraints $
\dot\gamma_D\cdot\gamma=0$ (conservation of $\Tr\rho=1$) and $
\dot\gamma_D\cdot H'=0$ (energy conservation). Thus, we assume
$\dot\gamma_D$  orthogonal  to $\gamma$ and $H'$. As a result it
also turn out to be  orthogonal to $\dot\gamma_H$.

To identify the direction of steepest entropy ascent, we follow a
simple geometrical construction based on the well-known but seldom
used standard geometrical notions reviewed in the Appendix, which
from here on we give for granted. Let $L(\gamma,H')$ denote the
real linear span of vectors $\gamma$ and $H'$. Denote by $S'_L$
the orthogonal projection of the entropy gradient vector $S'$ onto
$L$, and by $S'_{\perp L}$ its orthogonal complement, so that
\begin{equation}S' = S'_L + S'_{\perp L}\ .\end{equation} Assume for
simplicity that $\gamma$ and $H'$ are linearly independent (the
case in which they are not is easily covered as done in the
Appendix). Then, we may write
\begin{align}\label{S'gram} { S' }_{\bot  L(\gamma,H')}&= \fr{\lat
\begin{array}{ccc} S' & \gamma & H' \\ \\ S'
\cdot \gamma & \gamma \cdot \gamma  & H' \cdot \gamma
\\ \\ S' \cdot H' &
\gamma \cdot H'  & H' \cdot H'
\end{array} \rat}{\lat
\begin{array}{cc} \gamma \cdot \gamma & \gamma \cdot H' \\ \\
\gamma \cdot H' & H' \cdot H' \end{array} \rat} =\fr{\lat
\begin{array}{cc} (\Delta S)'  & (\Delta H)' \\  \\ (\Delta S)' \cdot (\Delta H)' &
 (\Delta H)' \cdot (\Delta H)'
\end{array} \rat}{
(\Delta H)'\cdot (\Delta H)' } \\ \nonumber\\&=  (\Delta
S)'-\frac{(\Delta S)' \cdot (\Delta H)'}{(\Delta H)' \cdot (\Delta
H)'}(\Delta H)'= (\Delta S)'-\frac{\cov{S}{H}}{\cov{H}{H}}(\Delta
H)' \ .
\end{align} Clearly, operator $S'_{\bot L}$ is the component of the
entropy gradient operator $S'$ orthogonal to both $\gamma$ and
$H'$. Using the above expression, it can be readily verified that
$S'_{\bot L}$ is orthogonal also to $\dot\gamma_H$
(\ref{gammadotH}). Moreover, $S'_{\bot L}\cdot S'_{\bot
L}=\cov{S}{S}-\cov{H}{S}^2/\cov{H}{H}$.

Therefore, we assume that $\dot\gamma_D$ is in the
\quot{direction} of $S'_{\bot L}$, and we let
\begin{equation}\label{gammadotD}\dot\gamma_D=
\frac{1}{4\tau_D}\frac{{ S' }_{\bot L(\gamma,H')}}{\sqrt{{ S'
}_{\bot L(\gamma,H')}\cdot{ S' }_{\bot L(\gamma,H')}}}\ ,
\end{equation} where $\tau_D$ is some positive constant or
functional of $\rho$ which we call the \quot{intrinsic
characteristic time of dissipation} also for the reasons to be
further discussed below. As done in our original work, for
simplicity of notation we also define the positive characteristic
time functional
\begin{equation}\label{tau}\tau = \tau_D\sqrt{{ S' }_{\bot
L(\gamma,H')}\cdot{ S' }_{\bot L(\gamma,H')}}/\Boltz \ \mbox{,
where $\Boltz$ is Boltzmann's constant.} \end{equation}

\subsection{\label{sec2-5}Dynamical Equation for the Density Operator of an
Isolated System}

When \quot{translated} in density operator formalism, our equation
of motion (\ref{gammadot}) with $\dot\gamma_D$ given by
(\ref{gammadotD}) may be written in the following compact form
\begin{equation}\label{rhodot}
\dot\rho=-\frac{i}{\hbar}[H,\rho]+
\frac{1}{2\tau_D\sqrt{\cov{M}{M}}}\{\Delta M,\rho\}\ \mbox{, where
} M=S-\frac{H}{\theta_H}  \mbox{ and } \theta_H
=\frac{\cov{H}{H}}{\cov{S}{H}} \ .
\end{equation}
Operator $M$, that we call the \quot{non-equilibrium Massieu
operator}, is a nonlinear function of $\rho$ not only through the
logarithmic dependence in operator $S$ but also through the
nonlinear functional $\theta_H$, which we may call
\quot{constant-energy nonequilibrium temperature}, because it will
smoothly approach the equilibrium temperature as the state will
approach at constant energy the maximal entropy stable equilibrium
state (see below). Notice also that any reference to the square
root operator $\gamma$ disappears, proving that its use is only
auxiliary to the given geometrical construction, as it is
auxiliary to the equivalent variational formulation given in
\cite{Gheorghiu} (see below). For this reason, the first
formulation \cite{thesis} avoided the explicit use of
$\sqrt{\rho}$.

\section{\label{sec3}Features of the proposed dynamics for a
single isolated particle}

Detailed proofs of the general features of Eq. (\ref{rhodot}) that
we outline in this section are given in Refs.
\cite{ref41,ref42,ref44,ArXiv1,ArXivTE}.

\subsection{\label{sec3-1}Conservation of Nonnegativity of the
Density Operator}

Eq. (\ref{rhodot}) generates a smooth continuous path in state
space and maintains at zero any initially zero eigenvalue of
$\rho$. As a result, no eigenvalue can cross zero and become
negative, neither in the future nor in the past, thus assuring
conservation of the nonnegativity of $\rho$ both in forward and
backward time.

\subsection{\label{sec3-2}Entropy Generation Rate}

The  rate of entropy change (more precisely, entropy
\quot{generation}, since  so far Eq. (\ref{rhodot}) is meant to be
a model for an isolated or an adiabatic system) takes the
following various explicit expressions
\begin{align}\label{sdot} \frac{d\langle
S\rangle}{dt}&=\frac{d(-\Boltz\Tr\rho \ln\rho)}{dt}=
\dot\gamma\cdot S'=\frac{1}{4\Boltz\tau}S'_{\bot L} \cdot S'_{\bot
L} = 4\Boltz\tau\,\dot\gamma_D\cdot\dot\gamma_D
\\ \label{sdot2}&=\frac{1}{\Boltz\tau}\cov{M}{M}=
\frac{1}{\Boltz\tau}\left(\cov{S}{S}-\frac{\cov{H}{H}}{\theta_H^2}\right)
=\frac{1}{\tau_D}\sqrt{\cov{M}{M}} \ ,
\end{align}
and is clearly positive semi-definite owing to the well-known
positive semi-definiteness of Gram determinants (Appendix) and
scalar product norms [see the last two of Eqs. (\ref{sdot})].

\subsection{\label{sec3-3}Characteristic Times and Time-Entropy Uncertainty Relation}

Because of the orthogonality between $\dot\gamma_D$ and
$\dot\gamma_H$, if at one time $[H,\rho]=0$ then $\rho(t)$
commutes with $H$ at all times. For such particular, \quot{purely
dissipative} trajectories, the Fisher-Rao metric takes on the
following interesting explicit expressions
\begin{align}d\ell &= 2\sqrt{\dot\gamma_D\cdot\dot\gamma_D}
\, dt =
 \sqrt{\frac{1}{\Boltz\tau}\frac{d\langle
S\rangle}{dt}}\, dt= \frac{1}{\Boltz\tau}\frac{d\langle
S\rangle}{d\ell}\, dt \label{FRS}\\ &=
\frac{\sqrt{\cov{M}{M}}}{\Boltz\tau}\, dt =
\frac{1}{\Boltz\tau}\sqrt{\cov{S}{S}-\frac{\cov{H}{H}}{\theta_H^2}}\,
dt=\frac{dt}{\tau_D} \ , \label{tauD}
\end{align}
where the last equality justifies our calling $\tau_D$ the natural
\quot{intrinsic dissipative time}. The last of Eqs. (\ref{FRS})
shows that the alternate characteristic time $\tau$ is directly
related to the rate $d\ell/dt$ at which the state operator
$\gamma$ follows the steepest entropy ascent trajectory, modulated
by the dimensionless entropy gradient $d(S/\Boltz)/d\ell$ along
the trajectory,
\begin{equation}
\frac{1}{\tau}=\frac{d\ell/dt}{d(S/\Boltz)/d\ell} \ ,
\end{equation}
so, we see that when time intervals are measured in units of the
\quot{local} (or instantaneous) value of $\tau$ and entropy in
units of $\Boltz$, the \quot{speed} along the steepest entropy
ascent trajectory (geodesic) coincides with the local slope of the
entropy surface along the trajectory,
\begin{equation}\frac{d \ell}{dt/\tau}= \frac{d (S/\Boltz)}{d\ell}
\ , \end{equation} which again justifies the interpretation of
$\tau$ as an intrinsic dynamical time. Finally, we notice that
$dS/d\ell$ equals half of the norm of the component of the entropy
gradient operator $S'$ orthogonal to the linear manifold
$L(\gamma,H')$ defined by the gradients of the constraints (here,
for simplicity, only normalization and energy conservation),
\begin{equation}d S/d\ell=\onehalf
\sqrt{{ S' }_{\bot  L(\gamma,H')}\cdot{ S' }_{\bot L(\gamma,H')}}
\ .
\end{equation}

A noteworthy result follows from Eq. (\ref{sdot2}) together with
the general inequality $\cov{S}{S}\ge \cov{M}{M}$ (Appendix). By
defining the characteristic time $\tau_S$ of the entropy
generation,
 we find the
following general time-entropy uncertainty relations
\begin{equation}\label{timeentropy1} \tau^2_{S}=\frac{\cov{S}{S}}{|d\langle
S \rangle/dt|^2}\ge \frac{\cov{M}{M}}{|d\langle S
\rangle/dt|^2}=\frac{(\Boltz\tau)^2}{\cov{M}{M}}=\tau_D^2 \quad
\Rightarrow \quad \cov{M}{M}\,\tau^2_{S}\ge
(\Boltz\tau)^2\end{equation} and also
\begin{equation}\label{timeentropy2} \cov{S}{S}\,\tau^2_{S}\ge
(\Boltz\tau)^2 \ ,\end{equation} which imply that the rate of
entropy generation cannot exceed the following bounds,
\begin{equation}d\langle S
\rangle/dt\le \sqrt{\cov{S}{S}}/\tau_D \le \cov{S}{S}/\Boltz\tau \
.
\end{equation}

\subsection{\label{sec3-4}Equilibrium States and Limit Cycles}

From Eq. (\ref{sdot}) we see that the rate of entropy generation
is zero (and the evolution is Schr\"odinger--von Neumann) if and
only if $\dot\gamma_D=0$, i.e., when $S'$ lies in $L(\gamma,H')$.
Then, the density operator may be written as
 \begin{equation}\label{nondissH} \rho
= \frac{B \exp(-H/\Boltz T)B}{\Tr[ B \exp(-H/\Boltz T)B]} \
\mbox{, for some } B= B^2 \ , \end{equation} where
$T=\theta_H=\cov{H}{H}/\cov{H}{S}=\theta_S=\cov{H}{S}/\cov{S}{S}
=\sqrt{\cov{H}{H}/\cov{S}{S}}$. We call these the
\quot{nondissipative states}. Proofs of the above and following
results are straightforward, and detailed in the original papers
\cite{ref41,ref42,ref44,Frontiers}.

Because  $\dot\gamma_D=0$, nondissipative states remain
nondissipative at all times, $-\infty<t<\infty$. Therefore they
obey unitary Hamiltonian evolution. If $[B,H]=0$ they are
equilibrium states. If $[B,H]\ne 0$ their unitary evolutions are
limit cycles of the dynamics, and $B(t) = U(t) B(0) U^{-1}(t)$
with $U(t)=\exp(-itH/\hbar)$. Limit cycles can be mixed, if $\Tr
B>1$, or pure, if $\Tr B=1$. The latter case coincides with the
usual Schr\"odinger dynamics of standard quantum mechanics. Except
for when $B=I$ (the identity operator) all these equilibrium
states and limit cycles are unstable (according to Lyapunov).

\subsection{\label{sec3-5}Dynamical Group, Not a Semi-Group}

As proved in the original papers and again in Refs.
\cite{PRE,Gheorghiu}, the solutions of our nonlinear dynamical
equation form a group, not a semi-group, of \quot{trajectories} in
density operator space. This is so because of a very important
feature of the dynamics, namely, that along every trajectory the
zero eigenvalues of $\rho$ are invariant, the range of $\rho$ and
its cardinality $\Tr P_{\Ran\rho}$ are invariant. However, the
nonzero eigenvalues of $\rho$ get smoothly rearranged towards
higher entropy distributions approaching  partially canonical
distributions of the form \ref{nondissH}. Because positive
eigenvalues of $\rho$ remain positive at all times, it follows
that every trajectory is unique and well-defined not only in
forward time but also backwards in time (an explicit proof of the
uniqueness feature is given in \cite{twolevel} for a two level
system, and is also discussed in general in \cite{Gheorghiu}).

We therefore have a \quot{strongly causal} group of dynamical
evolutions, with inverse defined everywhere, unique trajectories
through every state $\rho$, fully defined both forward and
backward in time, thus allowing full reconstruction of the past
from the present. It is an example of an \emph{invertible}
dynamics which nevertheless is  fully compatible with all
thermodynamics principles and in addition is largely irreversible
in that in forward time and for a system which is isolated (or
\quot{adiabatic} in the sense that the Hamiltonian operator may be
time dependent) entails and describes explicitly entropy
generation along the direction of maximal increase.

A remark about invertibility is in order. An often encountered,
misleading assertion is that  to be \quot{irreversible} a
dynamical evolution  must not have  an inverse. To justify the
theory of dynamical semi-groups, the misleading suggestion has
often been made that non-invertibility is an indispensable feature
for the description of thermodynamic irreversibility, so that the
equation of motion can be solved only forward in time, not
backwards, and \emph{causality} is thus retained only in a
\quot{weak form}: future states can be predicted from the present
state, but the past cannot be reconstructed from the present.

Our nonlinear dynamical group challenges this idea. The existence
of thermodynamic irreversibility is not incompatible with
causality in the strong sense: it does not necessarily forbid the
possibility to reconstruct the past from the present. If $\rho(0)$
and $\rho(t)$ are the states at times 0 and $t$, they are related
by the dynamical map $\rho(t)=\Lambda_t(\rho(0))$ through the
solution of the equation of motion for the time interval from 0 to
$t$ with initial condition $\rho(0)$. If the inverse map exists,
it points from the final state back to the initial state,
$\rho(0)=\Lambda^{-1}_t(\rho(t))$, and there is a one-to-one
correspondence between initial and final state. This is our case.
It shows that existence of the inverse map, does not rule out the
possibility that the functional which represents thermodynamic
entropy could be \emph{non-decreasing in forward time}. In Ref.
\cite{PRE} we present some numerical solutions which exemplify how
 any given distribution of eigenvalues belongs to a unique
smooth solution which among other features identifies (as
$t\rightarrow -\infty$) a lowest-entropy (not necessarily
zero-entropy) \quot{ancestral} or \quot{primordial} state.

\subsection{\label{sec3-6}Stability of Equilibrium States and the Second Law of
Thermodynamics}

Each partially canonical equilibrium density operator of the form
(\ref{nondissH}) maximizes the entropy when restricted to the
subset of density operators that share the same kernel. Such
states are equilibrium (i.e., time invariant if $H$ is time
invariant), but are unstable whenever at least one eigenvalue of
$\rho$ is equal to zero.  In fact, a minor perturbation which
changes the zero eigenvalue to an arbitrarily small nonzero value,
would proceed away towards a quite different equilibrium of higher
entropy. Instead, any trajectory with no null eigenvalues,
maintains such feature at all times (invariance of the cardinality
of the set of eigenvalues), and approaches in forward time the
unique, fully canonical, maximum-entropy density operator
compatible with the initial values of the constraints (remember
that for simplicity we are considering here only systems whose
approach to equilibrium is constrained only by energy conservation
and, of course, normalization).

Therefore, the only dynamically stable equilibrium states (again,
stable according to Lyapunov) are those given by Eq.
(\ref{nondissH}) with $B=I$.  There is only one such  canonical
density operator  for every value of the mean energy $\langle
H\rangle$, which through $\Tr H\rho=\langle H\rangle$ fixes the
temperature $T$ in (\ref{nondissH}). Existence and uniqueness of
stable equilibrium states for every value of the energy is the
essence of the Hatsopoulos-Keenan statement of the second law of
thermodynamics, which we may state as follows \cite{HG,GB2005,HK}:
\emph{Among all the states of a system that have a given value of
the energy and are compatible with a given set of values of the
amounts of constituents and the parameters of the  Hamiltonian,
there exists one and only one stable equilibrium state.} From this
statement of the second law, the Kelvin-Planck, the Clausius, and
the Carath\'{e}odory statements can all be shown to follow as
logical consequences (explicit proofs of this assertion can be
found in \cite[p.64-65 (Kelvin-Planck), p.133-136 (Clausius), and
p.121 (Carath\'{e}odory)]{GB2005}).

This statement of the second law brings out very clearly the
apparent conflict between mechanics and thermodynamics, a contrast
that for over a century has been perceived as paradoxical. In
fact, within mechanics, classical or quantum, the following
so-called \emph{minimum energy principle} applies: \emph{Among all
the states of a system that are compatible with a given set of
values of the amounts of constituents and  the parameters of the
Hamiltonian, there exists one and only one stable equilibrium
state, that of minimal energy}. Comparing this assertion with the
statement of the second law just reviewed, leads to a paradox if
we insist that the two theories of Nature contemplate the same set
of states. Indeed, for fixed amounts of constituents and
parameters of the Hamiltonian, mechanics asserts the existence of
a unique stable equilibrium state (that of minimal energy),
whereas thermodynamics asserts the existence of infinite stable
equilibrium states (one for every value the mean energy can take).

The paradox is  removed if we admit
  that the \quot{pure} states
contemplated by Quantum Mechanics are only a subset of those
contemplated by Thermodynamics. This resolving assumption was very
controversial when  Hatsopoulos and Gyftopoulos first introduced
it in  \cite{HG}. However today---more as a byproduct of the more
recent vast literature on quantum entanglement and quantum
information than as a result of thermodynamic reasoning---an
assumption to this effect is ever more often being included in the
postulates of quantum theory (compare for example the postulates
of quantum theory as stated, e.g., in the recent \cite{Ozawa2004}
with those stated in 1968 by Park and Margenau  \cite {PM1968}).
The discussions on the relations between this fundamental
assumption and thermodynamics is flourishing in the physics
literature (see \cite{Challenge,Alicki2004} and references
therein). Unfortunately, pioneering contributions such as
\cite{HG} are seldom acknowledged.

As mentioned in the introduction, the validity of the present
steepest-entropy-ascent or maximal-entropy-generation mathematical
formalism even outside of the original framework
 for which it was developed \cite{Nature,HG,ref41,ref42} has been
recognized and suggested by this author long ago, not only for
quantum dynamical phenomenological modeling \cite{Frontiers} but
also as a general tool for modeling relaxation and redistribution
of nonequilibrium probability or other positive-valued
distributions in a variety of fields \cite{ASME,Entropy}.

\subsection{\label{sec3-7}Variational Formulation}

In 2001, Gheorghiu-Svirschevski \cite{Gheorghiu} re-derived Eq.
(\ref{rhodot}) from a variational principle that in our notation
is
\begin{equation} {\rm max}\
\frac{d\langle S\rangle}{dt} \mbox{ subject  to }\frac{d\langle
H\rangle}{dt}=0,\ \frac{d\Tr\rho}{dt}=0 \mbox{, and  }
\dot\gamma_D\cdot\dot\gamma_D =c^2 ,
\end{equation}
where   the last constraint signifies that  we maximize the rate
of entropy generation at fixed norm of the operator
$\dot\gamma_D$, hence we are free to vary only its direction
($c^2$ is some real functional independent of $\dot\gamma_D$).
Introducing Lagrange multipliers,
\begin{equation}L = \dot\gamma_D\cdot S'-\lambda_1\, \dot\gamma_D\cdot \gamma
-\lambda_H\, \dot\gamma_D\cdot H' -
\lambda_\tau\,\dot\gamma_D\cdot\dot\gamma_D  \ ,
\end{equation} and maximizing $L$ with respect to $\dot\gamma_D$ yields
\begin{equation}\label{lagrange} S'-\lambda_1\,  \gamma
-\lambda_H\,  H' - 2\lambda_\tau\,\dot\gamma_D =0 \ ,
\end{equation}
where the multipliers must be determined by substitution in the
 constraint equations. It is easy to verify that our
expression of $\dot\gamma_D$ in Eqs. (\ref{S'gram}) and
(\ref{gammadotD}) yields the explicit solution of Eq.
(\ref{lagrange}). Using  (\ref{tauD}) we see that with
$c^2=1/4\tau_D^2$ we get exactly our quantum dynamical evolution
equation.

\subsection{\label{sec3-8}Onsager Reciprocal Relations even
Far from Equilibrium}

Any nonequilibrium $\rho$ can be written as
\begin{equation}\label{noneqrho} \rho =\frac{ B\exp(-\sum_j f_j
X_j)B}{\Tr B\exp(-\sum_j f_j X_j)} \ ,  \end{equation} where the
set $\{I,X_j\}$ spans the real space of hermitian operators on
$\Hil$, and $B=B^2$ is a projector (actually, $B=
P_{\Ran\rho}$).\footnote{To prove Eq. (\ref{noneqrho}), let
$P_{\Ker\rho}$ denote the projector onto the kernel of $\rho$,
i.e., the eigenspace belonging to the zero eigenvalue (if $\rho$
is non-singular, $P_\Ker\rho$ projects onto the null vector of
$\Hil$), and let $B(\Hil)$ denote the real space of the hermitian
operators on $\Hil$, equipped with the inner product $X\cdot
Y=\Tr(XY)$ and let $\{I,X_j\}$ be a basis for $B(\Hil)$. The
(nonnegative definite) operator $-\ln(\rho+P_{\Ker\rho})$ is well
defined for every $\rho$ and, since it belongs to $B(\Hil)$, we
may write it as $-\ln(\rho+P_{\Ker\rho})=f_0 I+\sum_j f_j X_j$.
Therefore, $\rho+P_{\Ker\rho}=e^{-f_0}\exp(-\sum_j f_j X_j)$.
Multiplying by $P_{\Ran\rho}$ and using the identities
$P_{\Ran\rho}P_{\Ker\rho}=0$ and $P_{\Ran\rho}\rho=\rho
P_{\Ran\rho}=\rho$, we obtain
$\rho=e^{-f_0}P_{\Ran\rho}\exp(-\sum_j f_j X_j)$. Finally, by
imposing $\Tr\rho=1$, we find
$e^{f_0}=\Tr[P_{\Ran\rho}\exp(-\sum_j f_j X_j)]$.} We may call the
set $\{X_j\}$ a \quot{quorum} of observables, because the
measurement of their mean values $\{\langle X_j\rangle\}$ fully
determines  the density operator. The empirical determination of a
quantum state has been recently called \quot{quantum tomography}.
In an almost forgotten seminal series of papers in 1970-1971, Park
and Band \cite{PB1970} devised elegant systematic rules to
construct such a quorum of observables.

Given such a quorum, we can write
 \begin{equation}   \langle X_j\rangle =\Tr (\rho X_j)\mbox{, }
 \langle S\rangle=\Boltz f_0 + \Boltz\sum_j f_j\, \langle X_j\rangle
 \mbox{, and } \displaystyle\Boltz f_j =
\left.\frac{\partial \langle S\rangle}{\partial
 \langle X_j\rangle}\right|_{ \langle X_{i\ne j}\rangle}  ,
\end{equation}  where $\Boltz f_j$ may be interpreted as the
\quot{generalized affinity} or force, conjugated with the
observable associated with operator $X_j$.

Let us focus on the dissipative term in our equation of motion
(\ref{rhodot}) and the rate of change it induces on the mean value
$\langle X_j\rangle$ of each quorum observable. We call it the
\quot{dissipative (part of the) rate of change} of observable
$X_j$,
\begin{equation} {\langle
\dot X_j\rangle_D} = \dot\gamma_D\cdot X'_j \mbox{ , with
}X'_j=2\gamma X_j \ .
\end{equation}
From the expressions we derived for $\dot\gamma_D$ in our steepest
entropy ascent dynamics, we find the following \emph{linear}
relations between dissipative rates and affinities
\begin{equation}
{\langle \dot X_i\rangle_D} =\sum_j f_j \, L_{ij}(\rho) \mbox{ , }
\end{equation}
where the coefficients are \emph{nonlinear} functionals of $\rho$
which form a symmetric, positive semi-definite Gram matrix $[
\{L_{i j}(\rho)\}]$,  \quot{generalized conductivity},
\begin{equation} L_{i j}(\rho)={\displaystyle \frac{1}{\tau}}
\frac{\left|
\begin{array}{cc} {\cov{X_i}{X_j}}
&{\cov{H}{ X_j }}\\ \\ {\cov{ X_i }{H}} &{\cov{H}{H}} \end{array}
\right| }{\cov{H}{H}}  = L_{ji}(\rho) \mbox{ , }
\end{equation}
As a result, the entropy generation rate may be written as a
quadratic form in the affinities
\begin{equation}
\frac{d\langle S\rangle}{dt} =\Boltz\sum_i\sum_j f_i f_j L_{i
j}(\rho)  \mbox{ . }
\end{equation}
When $[ \{L_{i j}(\rho)\}]$ is positive definite, we denote its
inverse, \quot{generalized resistance}, by $[ \{R_{i j}(\rho)\}]$,
\begin{equation} f_j=\sum_i R_{i j}(\rho){\langle \dot
X_i\rangle_D} \mbox{ , }
\end{equation} and the entropy generation rate may then be written also as a
quadratic form in the dissipative rates
\begin{equation} \frac{d\langle S\rangle}{dt} =\Boltz\sum_i\sum_j
L^{-1}_{ij}(\rho){\langle \dot X_i\rangle_D} {\langle \dot
X_j\rangle_D}  \mbox{ , }
\end{equation}
as well as as a sum of the dissipative rates of change of the
quorum observables each multiplied by its conjugated affinity
\begin{equation} \frac{d\langle S\rangle}{dt} =\Boltz\sum_i
f_i\,{\langle \dot X_i\rangle_D}  \mbox{ . }
\end{equation}

Notice that the parametrization of density operators given by Eq.
\ref{noneqrho} in terms of the real variables $f_j$,  implies that
we can write the entropy as $\langle S\rangle=\langle
S\rangle(f_1,\dots,f_{N^2-1})$ in view of the fact that our
dynamics conserves the cardinality of $\rho$ (and hence $\Tr B$ is
invariant).

\subsection{\label{Master}Nonlinear Master Equation for Energy
Level Occupation Probabilities}

When written for the $nm$-th matrix element of $\rho$ with respect
to an eigenbasis $\{|\epsilon_j\rangle\}$ of $H$, Eq.
(\ref{rhodot}) becomes
\begin{equation}\label{master1} \ddt{\rho_{nm}} =
 -\frac{i}{\hbar} \rho_{nm} ( E_n - E_m )+ \frac{1}{\Boltz\tau}
 \sum_r u_{nr}u^*_{mr}p_r\left(\Delta s_r- \frac{\Delta e_n+\Delta
 e_m}{2\theta_H}\right)  \mbox{ , }
\end{equation}
where $u_{jk}=\langle\epsilon_j|\eta_k\rangle$,
$\{|\eta_k\rangle\}$ is an eigenbasis of $\rho$, $p_k$'s its
eigenvalues, $\Delta s_k=s_k-\langle S\rangle$, $s_k=-\Boltz\ln
p_k$ if $p_k\ne 0$, $s_k=0$ if $p_k=0$, $e_j$ the eigenvalues of
$H$, $\Delta e_j=e_j-\langle H\rangle$, and $\theta_H$, $\Boltz$
and $\tau$ as already defined above. From (\ref{master1}) we see
that if at one instant of time, $[H,\rho]=0$ and we select a
common eigenbasis, then $u_{jk}=\delta_{jk}$ and $d\rho_{nm}/dt=0$
for $n\ne m$, which means that the condition $[H,\rho]=0$ holds
along the entire trajectory. In such special but nontrivial cases,
the eigenvalues of $\rho$ get redistributed according to the
nonlinear master equation
\begin{equation}\label{master2} \ddt{ p_n} = \frac{1}{\Boltz\tau}
 p_n\left(\Delta s_n-\frac{\Delta e_n}{\theta_H}\right) \mbox{ , }
\end{equation}
whose fundamental features are analyzed and numerically
exemplified  in Ref. \cite{PRE}, where we wrote it in the
following equivalent form
\begin{equation}
 \ddt{ p_n}=-\frac{1}{\tau}\left[p_n\ln
p_n+\alpha\,p_n+\beta\,e_np_n\right] \mbox{ , } \end{equation}
with the nonlinear functionals $\alpha$ and $\beta$ defined by
\begin{equation} \alpha= \frac{\sum_i e_ip_i\,\sum_j e_jp_j\ln
p_j- \sum_i p_i\ln p_i\,\sum_j e_j^2p_j}{ \sum_i e_i^2p_i
-\Big(\sum_i e_ip_i\Big)^2 } \mbox{, } \beta= \frac{\sum_i p_i\ln
p_i\,\sum_j e_jp_j-\sum_i e_ip_i\ln p_i}{ \sum_i e_i^2p_i
-\Big(\sum_i e_ip_i\Big)^2 }  \mbox{. }\end{equation}

In the usual statistical mechanics framework, the eigenvalues of
$\rho$ when $[\rho,H]=0$ are interpreted as \quot{occupation
probabilities}, meaning that the system is to be thought of as in
a particular, unknown, pure state; then, these
\quot{probabilities} are understood as an expression of the
uncertainty as of which pure state the system is actually
\quot{occupying}. In our original framework, instead, the
eigenvalues of $\rho$ when $[\rho,H]=0$ are interpreted as
\quot{degrees of energy load sharing} among the different modes
(eigenvectors of $H$) with which the system can internally
accommodate its mean energy. From this point of view, entropy
measures an overall degree of sharing between the available and
active modes (i.e., those with non-zero eigenvalues). Entropy
generation measures therefore the rate at which the spontaneous
internal dynamics redistributes energy among the available modes,
to achieve maximal sharing.

\section{\label{Composite}Extension of Equation (\ref{rhodot}) to
Composite Systems}

The nonlinear, dissipative term $\dot\gamma_D$ in Eq.
(\ref{gammadot}) provides a strong coupling between the energy
storage modes of the single-particle system, additional to the
coupling entailed by the linear, unitary term $\dot\gamma_H$
through the structure of the particle's Hamiltonian operator $H$.
Were we to apply Eq. (\ref{gammadot}) without modifications to a
system composed of two particles $A$ and $B$ (or to a more complex
composite system) the term $\dot\gamma_D$ would couple the
subsystems and make them exchange energy even in the absence of an
interaction term in the Hamiltonian $H$, thus violating both
separability and no-signaling criteria. Because of the
nonlinearity which is intrinsic in the steepest entropy ascent
construction, if the model equation of motion is to meet these
criteria, for a composite system, the structure of the
interactions and the internal constraints between subsystems must
be described not only through the Hamiltonian operator, but also
through the structure of the dynamical equation
itself.\footnote{It is noteworthy that if the proposed nonlinear
evolution law is supposed to be a fundamental law of nature (i.e.,
not just a phenomenological modeling tool), then one should
specify criteria for dividing a system into its separate
elementary constituents. A unitary Hamiltonian dynamics depends
only on the Hamiltonian operator $H$, regardless of  the level of
description, i.e., of whether we reach the given $H$ by
considering as elementary constituents the individual atoms, or
the individual electrons within an atom and the nucleus, or the
nucleons, or quark and gluons, or the electron field and quark
field, etc. For the given $H$ and a given mean value $\langle
H\rangle$ of the energy, also the unique stable equilibrium state,
$\rho_{\langle H\rangle}=\exp(-\beta_{\langle H\rangle}
H)/\Tr\exp(-\beta_{\langle H\rangle} H)$, is independent of the
level of description. But if   the relaxation to stable
equilibrium is  described by our nonlinear law, then the dynamics
depends strongly  on the assumed level of description, because
through  a unique internal relaxation time for each elementary
subsystem, this dynamics fully couples all the internal modes of
the elementary subsystem in a local effort to follow a path of
steepest ascent in the locally perceived value of the overall
entropy (see below).}

Suppose Alice and Bob, $A$ and $B$, are the two elementary
subsystems of an adiabatic system. Each subsystem is   a single
particle. Alice and Bob may be  \begin{align*} \mbox{interacting:
} &H\ne H_A\otimes I_B+I_A\otimes H_B ;\\ \mbox{noninteracting: }
&H=H_A\otimes I_B+I_A\otimes H_B  ;\\ \mbox{correlated/entangled:
}&S(\rho)\ne S(\rho_A)\otimes I_B+I_A\otimes S(\rho_B) ; \\
\mbox{uncorrelated: } &S(\rho)= S(\rho_A)\otimes I_B+I_A\otimes
S(\rho_B) ;
\end{align*}
where here $S(\rho)$ denotes as before the operator $-\Boltz
P_{\Ran \rho}\ln\rho$,  $\rho$ is the density operator of the
overall system, and $\rho_A$, $\rho_B$ the reduced local density
operators.

Our construction \cite{thesis,ref42} was designed so as to obtain
a dynamical system obeying the following separability and
no-signaling criteria \cite{MPLA}. \begin{itemize} \item For
permanently non-interacting subsystems $A$ and $B$, every
trajectory passing through a state in which the subsystems are in
independent states ($\rho=\rho_A\otimes\rho_B$, where
$\rho_A=\Tr_B\rho$ and $\rho_B=\Tr_A\rho$) must proceed through
independent states along the entire trajectory, i.e., when two
uncorrelated systems do not interact with each other, each must
evolve in time independently of the other.
\item If at some instant of time two subsystems $A$
and $B$, not necessarily non-interacting, are in independent
states, then the instantaneous rates of change of the subsystem's
entropies $-\Boltz\Tr (\rho_A\ln\rho_A)$ and $-\Boltz\Tr
(\rho_B\ln\rho_B)$ must both be nondecreasing in time. \item Two
non-interacting subsystems $A$ and $B$ initially in correlated
and/or entangled states  (possibly due to a previous interaction
that has then been turned off) should in general proceed in time
towards less correlated and entangled states (\emph{impossibility
of spontaneous creation of any kind of correlations}). \item When
subsystems $A$ and $B$ are not interacting, even if they are in
entangled or correlated states, it must be impossible that the
time dependence of any local observable of one subsystem be
influenced by any feature of the time evolution of the other
subsystem (\emph{no-signaling condition}).
\end{itemize}

Notice that we do not request  that existing entanglement and/or
correlations between $A$ and $B$ established by past interactions
should have no influence whatsoever on the time evolution of the
local observables of either $A$ or $B$. In particular, there is no
physical reason to request (as is often done) that two different
states $\rho$ and $\rho'$ such that $\rho'_A=\rho_A$ should evolve
with identical local dynamics (${\rm d}\rho'_A /{\rm d}t ={\rm
d}\rho_A /{\rm d}t$) whenever $A$ does not interact with $B$, even
if entanglement and/or correlations in state $\rho$ differ from
those in state $\rho'$. Rather, we see no reasons why the two
local evolutions could not be different until spontaneous
decoherence (if any) will have fully erased memory of the
entanglement and the correlations established by the past
interactions now turned off. In fact, this may be a possible
experimental scheme to detect spontaneous decoherence. In other
words, we will not assume that the local evolutions be necessarily
Markovian.

Compatibility with the predictions of quantum mechanics about the
generation of  entanglement between interacting subsystems that
emerge through the Schr\"odinger-von\,\,Neumann term
$-i[H,\rho]/\hbar$, requires that the dissipative term  may entail
spontaneous loss of entanglement and loss of correlations between
subsystems, but should not be able to create them.

To this end, we devised  \cite{thesis,ref42} a construction which
hinges on the definitions of the following \quot{locally perceived
energy} and \quot{locally perceived entropy} operators, as a
result of which our composite dynamics  implements the ansatz of
\quot{steepest locally perceived entropy ascent},
\begin{eqnarray}
(\Delta H)^A=\Tr_B[(I_A\otimes \rho_B) \Delta
H]\quad&\mbox{and}&\quad (\Delta H)^B=\Tr_A[(\rho_A\otimes I_B)
\Delta H] \ , \\ (\Delta S)^A=\Tr_B[(I_A\otimes \rho_B) \Delta S]
\quad&\mbox{and}&\quad (\Delta S)^B=\Tr_A[(\rho_A\otimes I_B)
\Delta S] \ .
\end{eqnarray}

A geometrical construction (details in \cite{thesis,ref42,ArXiv1})
analogous to that outlined in Section \ref{sec2} for a single
particle, leads us to a composite-system
steepest-locally-perceived-entropy-ascent dynamics with the form
\begin{equation}{\displaystyle \ddt{\rho} =
- \frac{i}{\hbar}}[H,\rho] + {\displaystyle  \frac{1}{
2\Boltz\tau_A} }\{(\Delta M)^A,\rho_A\}\Otimes\, \rho_B+
{\displaystyle \frac{1}{ 2\Boltz\tau_B} }
\rho_A\,\Otimes\,\{(\Delta M)^B,\rho_B\} \ ,\end{equation}  where
$\tau_A$, $\tau_B$ are local characteristic times and, for
$J=A,B$,
\begin{equation}
 (\Delta M)^J= (\Delta S)^J-(\Delta H)^J/\theta_{HJ} \mbox{ with }
\theta_{HJ}= \covJ{H}{H}/\covJ{S}{H} \ .
\end{equation}
Each local dissipative term separately \quot{conserves} the
overall system's mean energy $\langle H\rangle=\Tr(\rho H)$. Each
subsystem's contribution to the overall system's rate of entropy
change is positive semidefinite
\begin{equation}
 \ddt{\langle S\rangle}=\frac{1}{ \Boltz\tau_A}\covA{M}{M}
 +\frac{1}{ \Boltz\tau_B}\covB{M}{M} \ .
\end{equation}

 If Alice and Bob
interact, it is only the Hamiltonian term in the evolution
equation which during the interaction builds up correlations. Once
generated, these correlations survive even after $A$ and $B$
separate, even if the loose touch completely. When that happens,
$A$ and $B$ remain correlated but begin to evolve independently of
one another. This is reflected in the local structure of our
equation and in particular of operators $(\Delta S)^A$, $(\Delta
S)^B$, $(\Delta H)^A$, $(\Delta H)^B$.

Despite the nonlinearity, the equation prevents no-signaling
violations, in that it satisfies the following \emph{strong
separability} conditions. Namely, denoting by
$\dot\rho_{AB}(\rho,H)$ the rhs of the equation, it is easy to
show that, for any $\rho$ and any $H_A$, $H_B$,
\begin{eqnarray} &&\Tr_B [\dot\rho_{AB}(\rho,H_A\Otimes
I_B\!+\!I_A\Otimes H_B)] =f_A \left((\Delta S)^A, \! H_A\right) \
, \label{strongseparabilityA}\\ & &\Tr_A
[\dot\rho_{AB}(\rho,H_A\Otimes I_B\!+\!I_A\Otimes
H_B)]=f_B\left((\Delta S)^B, \! H_B\right) \ ,
\label{strongseparabilityB}\end{eqnarray} Conditions
(\ref{strongseparabilityA}) and (\ref{strongseparabilityB}), when
restricted to uncorrelated states, $\rho= \rho_A\Otimes\rho_B$,
define the conditions of \emph{weak separability}, which of course
are a corollary of strong separability.

However, existing correlations do influence the local evolutions,
which therefore are not Markovian in that they do not depend only
on the respective local (reduced) states $\rho_A$ and $\rho_B$.

\section{\label{Further}Additional Phenomenological Modeling Equations
for the Density Operator of a Coupled System}

In this section, we return to the problem of describing the
effective interaction between a system and a reservoir. But
instead of starting from Hamiltonian dynamics and adopting
suitable approximations so as to arrive at the KSGL equation as
discussed in Section \ref{sec1-1}, we take a fully
phenomenological approach. Using our geometrical construction, it
is easy to \quot{design} dynamical equations that exhibit
dynamical features that we expect from typically thermodynamical
energy balance and entropy balance considerations. So, we may
obtain a variety of similar dynamical equations that, though
perhaps not as fundamental, may nevertheless be very useful in the
phenomenological description of nonequilibrium phenomena of
non-isolated systems.

For example, recent major advances in micro- and
nano-technological applications, often call for a detailed
description of the time evolution of non-equilibrium states that
are far from thermodynamics equilibrium and cannot be described by
partially canonical entropy density operators. In such far
nonequilibrium regime, the assumption of linearity underlying the
standard theory of irreversible processes may be cease to hold.
Yet  we may need to describe the simultaneous energy and entropy
exchange which occurs between our quantum system  and, say, a
thermal reservoir at temperature $T_Q$, whereby the ratio of the
energy to the entropy exchanged is equal to $T_Q$.

Thus, we will consider additional terms of the form
\begin{equation}\label{model}
\dot\rho=\cdots +\frac{1}{2\tau_G\sqrt{\cov{G}{G}}}\{\Delta
G,\rho\}\ \mbox{, with } G=S-\frac{H}{\theta} \ ,
\end{equation}
where the dots represent  other terms as in (\ref{rhodot}) or even
in the KGSL Eq. (\ref{Lindblad}),  $G$ is another \quot{
non-equilibrium Massieu operator} that depends on the choice of
 functional $\theta$ (see below), and $\tau_G$ is the
characteristic time of decrease of $\langle G\rangle$. For
$\theta=\theta_H$, $G$ coincides with operator $M$ in Eq.
(\ref{rhodot}). Notice that if $\theta$ is chosen to be a
constant, then $\tau_G d\langle G\rangle/dt=-\sqrt{\cov{G}{G}}< 0$
except for $\gamma G'=0$, that is, $\gamma S'=\gamma H'/\theta$ or
$\cov{H}{H}=\theta^2\cov{S}{S}$ which occurs only for states of
the form (\ref{nondissH}) with $T=\theta$.

Eq. (\ref{model}) generates energy and entropy rates of change
according to
\begin{equation}\label{rates} \frac{d\langle H \rangle
}{dt}=\frac{\cov{H}{H}}{\tau_G\sqrt{\cov{G}{G}}}
\left(\frac{1}{\theta_H}-\frac{1}{\theta}\right) \mbox{ and }
\frac{d\langle S \rangle
}{dt}=\frac{\cov{S}{S}}{\tau_G\sqrt{\cov{G}{G}}}
\left(1-\frac{\theta_S}{\theta}\right) \ ,
\end{equation}
where $\theta_H=\cov{H}{H}/\cov{H}{S}$ and
$\theta_S=\cov{H}{S}/\cov{S}{S}$ as defined before. Note that in
general $\theta_H\ge\theta_S$ with strict equality only at states
of form (\ref{nondissH}), and $\theta_H\theta_S\ge 0$ which means
they always have the same sign (that of $\cov{H}{S}$).

Alternatively, we may also consider  equations of the form
\begin{equation}\label{modelHelmholtz}
\dot\rho=\cdots -\frac{1}{2\tau_F\sqrt{\cov{F}{F}}}\{\Delta
F,\rho\}\ \mbox{, with } F=H-\theta S \ ,
\end{equation}
where  $F$ is a \quot{non-equilibrium Helmholtz free energy
operator} that depends on the choice of  functional $\theta$. In
this case,
\begin{equation}\label{ratesHelmholtz} \frac{d\langle H \rangle
}{dt}=-\frac{\cov{H}{H}}{\tau_F\sqrt{\cov{F}{F}}}
\left(1-\frac{\theta}{\theta_H}\right) \quad\mbox{and}\quad
\frac{d\langle S \rangle
}{dt}=-\frac{(\theta_S-\theta)\cov{S}{S}}{\tau_F\sqrt{\cov{F}{F}}}
\ .
\end{equation}
Eqs. (\ref{modelHelmholtz}) and  (\ref{ratesHelmholtz}) are
related to (\ref{model}) and (\ref{rates}) by the fact that
$\cov{F}{F}/\cov{G}{G}=\cov{H}{H}/\cov{S}{S}=\theta_H\theta_S$.

\subsection{Smooth Isoentropic Extraction of the Adiabatic Availability}

When a system is in a nonequilibrium state, we call
\quot{adiabatic availability} the largest amount of energy that
can be extracted in the form of work without leaving any other
effects external to the system and without changing the system's
Hamiltonian operator.\footnote{Even if the Hamiltonian operator is
a function $H(\lambda)$ of some controllable parameters, so that
the state of the system is given by $(rho,\lambda)$, the adiabatic
availability is the largest work that can be extracted with no net
changes in $\lambda$.} It is given by $\langle\Psi\rangle=\langle
H\rangle- \langle H\rangle_s$ where $\langle H\rangle=\Tr[H\rho]$
the mean energy of the nonequilibrium state and $\langle
H\rangle_s=\Tr[H\rho_s(H)]$ the mean energy of the unique stable
equilibrium state $\rho_s(H)=\exp(-\beta_s H)/\Tr\exp(-\beta_s H)$
that has the same entropy as the given state $\rho$, i.e., such
that $\langle S\rangle- \langle S\rangle_s$. As thoroughly
discussed in \cite{GB2005} in general terms, and in
\cite{HG,Allaverd} in the quantum framework, the adiabatic
availability cannot in general be completely extracted by means of
a unitary evolution, owing to the fact that a unitary process
cannot change the eigenvalues of $\rho$. Instead, a process is
required that while maintaining the entropy invariant, smoothly
modifies the eigenvalues of $\rho$ until they become canonically
distributed. At the end of this isoentropic change of state,
$\rho$ has the form (\ref{nondissH}) with $B=I$.

Whereas finding a practical way to control and interact with the
system's dynamics so as to extract its adiabatic availability from
an arbitrary initial state may be a very hard problem, our
geometrical construction makes it straightforward to design a
dynamical equation that describes phenomenologically such an
extraction, along a steepest-energy-descent trajectory at constant
entropy. It suffices to take $\dot\gamma$ proportional to
$-H'_{\bot L(\gamma,S')}$, i.e., the component the energy gradient
$H'$ orthogonal to both $\gamma$ and $S'$,
\begin{equation}\label{H'gram} { H' }_{\bot  L(\gamma,S')}=\fr{\lat
\begin{array}{cc} (\Delta H)'  & (\Delta S)' \\
 \\ (\Delta H)' \cdot (\Delta S)' &
 (\Delta S)' \cdot (\Delta S)'
\end{array} \rat}{
(\Delta S)'\cdot (\Delta S)' } = (\Delta
H)'-\frac{\cov{H}{S}}{\cov{S}{S}}(\Delta S)' \ .
\end{equation}
In terms of Eq. (\ref{modelHelmholtz}) for the density operator,
we may describe this by choosing
\begin{equation} \theta=\theta_S =\frac{\cov{H}{S}}{\cov{S}{S}} \ ,\quad
 F=F_\Psi=H-\theta_S S \quad\mbox{and}\quad \tau_F=\tau_{F_{\Psi}}
 \ ,
\end{equation}
therefore we may call $\theta_S$ the \quot{constant-entropy
nonequilibrium temperature} and $F_\Psi$ the
\quot{constant-entropy nonequilibrium Helmholtz free energy
operator}. With this choice of $\theta$ in Eq.
(\ref{modelHelmholtz}), the entropy remains constant while
$\theta_S$ and $\theta_H$ smoothly approach the temperature of a
canonical or partially canonical final state of lowest energy for
the given entropy. Notice that the rate of energy change $d\langle
H \rangle /dt$ is negative semidefinite (even for states with
negative $\theta_H$). Because the zero eigenvalues are time
invariant here like for Eq. (\ref{rhodot}), this term will extract
the full adiabatic availability only if the initial $\rho$ in
non-singular.

\subsection{Smooth Extraction of the Available Energy with Respect to a
Reservoir}

When a system is in a nonequilibrium state or in any state not of
mutual equilibrium with a given reservoir with temperature $T_R$,
we call \quot{available energy with respect to a reservoir with
temperature $T_R$} the largest amount of energy that can be
extracted in the form of work without any other effects external
to the combination of the system and the reservoir. It is given by
$\langle\Omega^R\rangle=\langle H\rangle- \langle H\rangle_R-
T_R\,(\langle S\rangle- \langle S\rangle_R)$ where $\langle
H\rangle_R$ and $\langle S\rangle_R$ are the energy and the
entropy of the unique stable equilibrium state
$\rho_R=\exp(-H/\Boltz T_R)/\Tr\exp(-H/\Boltz T_R)$  with
temperature $T_R$. Again, its definition is  discussed in
\cite{GB2005} in general terms, and in \cite{HG} in the quantum
framework.

We can design a dynamical equation that generates a trajectory
along a smooth descent in  available energy by taking Eq.
(\ref{modelHelmholtz}) with \begin{equation} \theta=T_R \ ,\quad
F=F_{\Omega^R}=H-T_R S \quad\mbox{and}\quad
\tau_F=\tau_{F_{\Omega^R}} \ ,
\end{equation}
where  $T_R$ is the constant temperature of the reservoir and
$F_{\Omega^R}$ is yet another \quot{ nonequilibrium Helmholtz free
energy}. With this choice of $\theta$, the signs of the energy and
entropy rates (\ref{ratesHelmholtz}) depend on those of
$1-T_R/\theta_R$ and $\theta_S-T_R$. This is a model of a
reversible weight process for the  system-reservoir composite
\cite{GB2005} where, by the energy and entropy balance equations,
$T_R\,d\langle S\rangle/dt$ and $d\langle S\rangle/dt$ equal
respectively the net rates of energy and entropy exchange (from
the reservoir to the system, if positive, from the system to the
reservoir, if negative). Therefore, the remaining power,
$-d\langle H\rangle/dt+T_Rd\langle S\rangle/dt= -d\langle
\Omega^R\rangle/dt$, is the rate of energy extraction in the form
of work. From Eqs. (\ref{ratesHelmholtz}) we see that the energy
and entropy rates are both zero only at state $\rho_R$, where
$\theta_H=\theta_S=T_R$.

\subsection{Nonequilibrium Heat Interaction}

As a final example,  we consider the model of an interaction
between our system in a nonequilibrium state $\rho$ and some
reservoir (heat bath) at $T_Q$, whereby  the ratio of the energy
and the entropy exchange rates is equal to $T_Q$. The usual
definition of a heat interaction at $T_Q$ (see \cite{GB2005} for a
rigorous definition) requires both interacting bodies to be in
states very close to their respective stable equilibrium states
with temperature $T_Q$, because only then the ratio of energy to
entropy exchanged is equal to $T_Q$. Therefore, the interaction we
are modeling here is an extension of the standard notion to when
one of the interacting systems  is far from thermodynamic
equilibrium (where temperature is not defined). It is easy to
verify that by taking Eq. (\ref{modelHelmholtz}) with
\begin{equation}
\theta=\theta_Q=\theta_S\frac{\theta_H-T_Q}{\theta_S- T_Q} \
,\quad F=F_{T_Q}=H-\theta_Q S \quad\mbox{and}\quad
\tau_F=\tau_{F_{T_Q}} \ ,
\end{equation}
we obtain a smooth trajectory where at all times $d\langle H
\rangle /dt =T_Q\,d\langle S \rangle /dt$.

\section{\label{Conclusions}Conclusions}

In this paper we discuss the geometrical construction and the main
mathematical features of the
maximum-entropy-production/steepest-entropy-ascent nonlinear
evolution equation proposed long ago by this author in the
framework of a fully quantum theory of irreversibility and
thermodynamics for a single isolated or adiabatic particle, qubit,
or qudit. The same mathematics has been recently rediscovered by
other authors, with various physical interpretations.

The nonlinear equation generates a dynamical group, not just a
semigroup like for KSGL dynamics. It provides a deterministic
description of irreversible conservative relaxation towards
equilibrium from an arbitrary initial density operator. It
satisfies a very restrictive stability requirement equivalent to
the Hatsopoulos-Keenan statement of the second law of
thermodynamics which therefore emerges as a general theorem of the
dynamics. It has smooth unique solutions both forward and
backwards in time.  Except for fully characterized families of
limit cycles and of equilibrium states the entropy functional is
strictly increasing in forward time and strictly decreasing in
backward time. Viewed as a model of the relaxation to equilibrium
of an isolated single particle system, this dynamics entails
thermodynamic irreversibility at the single particle level.

For a multipartite isolated or adiabatic system, we introduce a
nonlinear projection defining local operators that we interpret as
``local perceptions'' of the overall system's energy and entropy.
Each component particle contributes an independent local tendency
along the direction of steepest increase of the locally perceived
entropy at constant locally perceived energy. It conserves both
the locally-perceived energies and the overall energy, and meets
strong separability and non-signaling conditions, even though the
local evolutions are not independent of existing correlations.

In addition, we also show how the geometrical construction can
readily lead to a variety of thermodynamically relevant models,
such as the phenomenological descriptions of nonunitary
isoentropic evolutions achieving  full extraction of a system's
adiabatic availability or available energy with respect to a
reservoir, or the phenomenological descriptions of a nonunitary
nonequilibrium heat interaction.

\section*{Acknowledgements}
The author is indebted to a Referee for help in clarifying the
presentation and for suggesting comments that have been adopted
almost verbatim. This paper was written during a visit to MIT
under the UniBS--MIT-MechE faculty exchange Program co-sponsored
by the CARIPLO Foundation, Italy under grant 2008-2290.

\section*{Appendix. Orthogonal Decomposition of a Vector
with respect to a Linear Manifold}

In the paper, we make extensive use of the notation and relations
discussed in this appendix (see \cite{ASME,Entropy} for a general
but non-quantal  context).

Given a set of vectors $ \g{g}_0 , \, \g{g}_1 , \, \ldots , \,
\g{g}_n $, the symbol \begin{equation} L \lt \g{g}_0 , \, \g{g}_1
, \, \ldots , \, \g{g}_n \rt \label{A11} \end{equation} \noindent
will denote their linear span, i.e., the linear manifold
containing all the vectors that are (real) linear combinations of
$ \g{g}_0 , \, \g{g}_1 , \, \ldots , \, \g{g}_n $. Given another
vector $ \g{b} $, the symbol \begin{equation} { \g{b} }_{L \lt
\g{g}_0 , \, \g{g}_1 , \, \ldots , \, \g{g}_n \rt} \label{A12}
\end{equation} \noindent will denote the orthogonal projection of
$ \g{b} $ onto the linear manifold $ L \lt \g{g}_0 , \, \g{g}_1 ,
\, \ldots , \, \g{g}_n \rt $, namely, the unique vector in $ L \lt
\g{g}_0 , \, \g{g}_1 , \, \ldots , \, \g{g}_n \rt $ such that its
dot product with any other vector $ \g{g} $ in $ L \lt \g{g}_0 ,
\, \g{g}_1 , \, \ldots , \, \g{g}_n \rt $ equals the dot product
of $ \g{b} $ with $ \g{g} $, i.e., \begin{equation} \g{g} \cdot {
\g{b} }_{L \lt \g{g}_0 , \, \g{g}_1 , \, \ldots , \, \g{g}_n \rt}
= \g{g} \cdot \g{b} \label{A13} \end{equation} \noindent for every
$ \g{g} $ in $ L \lt \g{g}_0 , \, \g{g}_1 , \, \ldots , \, \g{g}_n
\rt $.

In terms of a set of linearly independent vectors $ \g{h}_1 , \,
\ldots , \, \g{h}_r $ spanning the manifold $ L \lt \g{g}_0 , \,
\g{g}_1 , \, \ldots , \, \g{g}_n \rt $, where clearly $ r \leq n
$, we can write two equivalent explicit expressions for the
projection $ { \lt \g{b} \rt }_{L \lt \g{g}_0 , \, \g{g}_1 , \,
\ldots , \, \g{g}_n \rt} $ of vector $ \g{b} $ onto $ L \lt
\g{g}_0 , \, \g{g}_1 , \, \ldots , \, \g{g}_n \rt $. The first is
\begin{equation} {  \g{b}  }_{L \lt \g{g}_0 , \, \g{g}_1 , \, \ldots , \,
\g{g}_n \rt} = \sum_{k = 1}^r \sum_{m = 1}^r \lt \g{b} \cdot
\g{h}_k \rt { \lqu M { \lt \g{h}_1 , \, \ldots , \, \g{h}_r \rt
}^{- 1} \rqu }_{k \, m} \g{h}_m \ ,\label{A14} \end{equation}
\noindent where $ M { \lt \g{h}_1 , \, \ldots , \, \g{h}_r \rt
}^{- 1} $ is the inverse of the Gram matrix \begin{equation} M \lt
\g{h}_1 , \, \ldots , \, \g{h}_r \rt = \lqu
\begin{array}{ccc} \g{h}_1 \cdot \g{h}_1 & \cdots & \g{h}_r \cdot
\g{h}_1 \\ \vdots & \ddots & \vdots \\ \g{h}_1 \cdot \g{h}_r &
\cdots & \g{h}_r \cdot \g{h}_r \end{array} \rqu \ .\label{A15}
\end{equation} \noindent The second expression is a ratio of two
determinants \begin{equation} { \g{b}  }_{L \lt \g{g}_0 , \,
\g{g}_1 , \, \ldots , \, \g{g}_n \rt} = - \, \fr{\lat
\begin{array}{cccc} 0 & \g{h}_1 & \cdots & \g{h}_r \\ \g{b} \cdot
\g{h}_1 & \g{h}_1 \cdot \g{h}_1 & \cdots & \g{h}_r \cdot \g{h}_1
\\ \vdots & \vdots & \ddots & \vdots \\ \g{b} \cdot \g{h}_r &
\g{h}_1 \cdot \g{h}_r & \cdots & \g{h}_r \cdot \g{h}_r \end{array}
\rat}{\lat
\begin{array}{ccc} \g{h}_1 \cdot \g{h}_1 & \cdots & \g{h}_r \cdot
\g{h}_1 \\ \vdots & \ddots & \vdots \\ \g{h}_1 \cdot \g{h}_r &
\cdots & \g{h}_r \cdot \g{h}_r
\end{array} \rat} \ , \label{A16} \end{equation} \noindent where the determinant
at the denominator, also given by $\det M  \lt \g{h}_1 , \, \ldots
, \, \g{h}_r \rt$,  is always strictly positive because the
vectors $ \g{h}_1 , \, \ldots , \, \g{h}_r $ are linearly
independent.

In the paper, our rate equations are expressed in terms of vectors
of the form \begin{equation} { \g{b} }_{\bot  L \lt \g{g}_0 , \,
\g{g}_1 , \, \ldots , \, \g{g}_n \rt}=\g{b} - { \g{b} }_{L \lt
\g{g}_0 , \, \g{g}_1 , \, \ldots , \, \g{g}_n \rt} = \fr{\lat
\begin{array}{cccc} \g{b} & \g{h}_1 & \cdots & \g{h}_r \\ \g{b}
\cdot \g{h}_1 & \g{h}_1 \cdot \g{h}_1 & \cdots & \g{h}_r \cdot
\g{h}_1 \\ \vdots & \vdots & \ddots & \vdots \\ \g{b} \cdot
\g{h}_r & \g{h}_1 \cdot \g{h}_r & \cdots & \g{h}_r \cdot \g{h}_r
\end{array} \rat}{\lat
\begin{array}{ccc} \g{h}_1 \cdot \g{h}_1 & \cdots & \g{h}_r \cdot
\g{h}_1 \\ \vdots & \ddots & \vdots \\ \g{h}_1 \cdot \g{h}_r &
\cdots & \g{h}_r \cdot \g{h}_r \end{array} \rat} \ ,\label{A17}
\end{equation} \noindent where in writing Equation \ref{A17} we
make use of Equation \ref{A16}. The vector $  {  \g{b} }_{\bot L
\lt \g{g}_0 , \, \g{g}_1 , \, \ldots , \, \g{g}_n \rt} $ is
orthogonal to manifold $ L \lt \g{g}_0 , \, \g{g}_1 , \, \ldots ,
\, \g{g}_n \rt $; indeed, the vector represented by Equation
\ref{A17} has the relevant property \begin{equation} \g{g}_k \cdot
{  \g{b} }_{\bot L \lt \g{g}_0 , \, \g{g}_1 , \, \ldots , \,
\g{g}_n \rt} = 0 \qquad k = 0 , \, 1 , \, \ldots , \, n
\label{A18} \end{equation} \noindent which follows directly from
Relation \ref{A13}, and hence the relation \begin{equation} \g{b}=
{  \g{b} }_{L \lt \g{g}_0 , \, \g{g}_1 , \, \ldots , \, \g{g}_n
\rt}+ { \g{b} }_{\bot L \lt \g{g}_0 , \, \g{g}_1 , \, \ldots , \,
\g{g}_n \rt}\end{equation} represents the unique orthogonal
decomposition of vector $ \g{b}$ with respect to manifold $ L \lt
\g{g}_0 , \, \g{g}_1 , \, \ldots , \, \g{g}_n \rt $.

 Moreover, we have the
other obvious, but relevant properties \begin{equation} \g{b}
\cdot  {  \g{b} }_{\bot L \lt \g{g}_0 , \, \g{g}_1 , \, \ldots ,
\, \g{g}_n \rt} = { \g{b} }_{\bot L \lt \g{g}_0 , \, \g{g}_1 , \,
\ldots , \, \g{g}_n \rt} \cdot  { \g{b} }_{\bot L \lt \g{g}_0 , \,
\g{g}_1 , \, \ldots , \, \g{g}_n \rt} \geq 0 \ ,\label{A19}
\end{equation} \noindent where the strict inequality applies
whenever $ \g{b} $ is not in $ L \lt \g{g}_0 , \, \g{g}_1 , \,
\ldots , \, \g{g}_n \rt $, and for any $\g{a}$ and $\g{b}$
\begin{equation} \g{a} \cdot  {  \g{b} }_{\bot L \lt \g{g}_0 , \,
\g{g}_1 , \, \ldots , \, \g{g}_n \rt} = { \g{a} }_{\bot L \lt
\g{g}_0 , \, \g{g}_1 , \, \ldots , \, \g{g}_n \rt} \cdot  { \g{b}
}_{\bot L \lt \g{g}_0 , \, \g{g}_1 , \, \ldots , \, \g{g}_n \rt}
\geq 0 \ .\label{A20} \end{equation}

An important formula which derives from Eq. \ref{A17}  and the
usual properties of determinants, is \begin{equation} { \g{b}
}_{\bot L \lt \g{g}_0 , \, \g{g}_1 , \, \ldots , \, \g{g}_n \rt}
\cdot  { \g{b} }_{\bot L \lt \g{g}_0 , \, \g{g}_1 , \, \ldots , \,
\g{g}_n \rt} =\frac{\det M(\g{b},\g{h}_1,\dots,\g{h}_r)}{\det
M(\g{h}_1,\dots,\g{h}_r)} \ .\end{equation} Moreover,  choosing
the set of linearly independent vectors $ \g{h}_1 , \, \ldots , \,
\g{h}_r $ so that $ \g{h}_r=2\gamma$ is the gradient of the
normalization constraint ($\Tr \rho=\gamma\cdot\g{h}_r/2$), and
defining the \quot{mean} functionals $\langle H_j\rangle=\Tr\rho
H_j =\g{h}_j\cdot\g{h}_r/4$ [in the paper, we assume $r=2$, with
$H_1=H$, the Hamiltonian operator, and $H_2=I$ the identity, so
that $\g{h}_1=H'=2\gamma H$ and $\langle H_1\rangle=\Tr\rho H =
\gamma\cdot H'/2 = (2\gamma )\cdot (2\gamma H)/4=
\g{h}_1\cdot\g{h}_2/4$], $\langle B\rangle=\g{b}\cdot\g{h}_r/4$
[in the paper, $B=S$, $\g{b}=S'$], and the \quot{deviation}
vectors $\Delta \g{h}_j =(\g{h}_j-\g{h}_r \langle H_j\rangle )/2
$, $\Delta \g{b} =(\g{b}-\g{h}_r \langle B\rangle )/2 $, it is
easy to show that \begin{equation} \frac{\det
M(\g{b},\g{h}_1,\dots,\g{h}_n)}{\det M(\g{h}_1,\dots,\g{h}_n)}=
\frac{\det
M(\Delta\g{b},\Delta\g{h}_1,\dots,\Delta\g{h}_{n-1})}{\det
M(\Delta\g{h}_1,\dots,\Delta\g{h}_{n-1})} \ .
\label{last}\end{equation}

As a final remark, we write the following generalized form of the
Cauchy-Schwarz inequality \begin{equation} \det
M(\Delta\g{b},\Delta\g{h}_1,\dots,\Delta\g{h}_{n-1}) \le \det
M(\Delta\g{h}_1,\dots,\Delta\g{h}_{n-1})\,
\Delta\g{b}\cdot\Delta\g{b} \ . \label{Schwarz}\end{equation}

\end{document}